\documentclass[12pt]{article}

\usepackage{scicite}

\usepackage{times}
\usepackage{amsmath}
\usepackage{amssymb}

\usepackage{graphics}
\usepackage{epsfig}
\usepackage{color}
\usepackage{amsmath}

\newenvironment{sciabstract}{%
\begin{quote} \bf}
{\end{quote}}

\renewcommand{\thefigure}{\textbf{\arabic{figure}}}

\newcommand{\Psib}{\mbox{\boldmath$\Psi$}}

\newcounter{lastnote}

\title{Intrusion in heterogeneous materials:  Simple global rules from complex micro-mechanics}

\author
{Hesam Askari,$^{1}$ Ken Kamrin,$^{1\ast}$\\
\\
\normalsize{$^{1}$Massachusetts Institute of Technology,}\\
\normalsize{77 Massachusetts Ave., Cambridge, MA 02139, USA}\\
\\
\normalsize{$^\ast$ E-mail:  kkamrin@mit.edu}
}

\date{}

\begin{document} 


\baselineskip24pt


\maketitle


\begin{sciabstract}

The interaction of intruding objects with deformable materials is a common phenomenon, arising in impact and penetration problems, animal and vehicle locomotion, and various geo-space applications.  The dynamics of arbitrary intruders can be simplified using Resistive Force Theory (RFT), an empirical framework originally used for fluids but works surprisingly well, better in fact, in granular materials.  That such a simple model describes behavior in dry grains, a complex nonlinear material, has invigorated a search to determine the underlying mechanism of RFT.   We have discovered that a straightforward friction-based continuum model generates RFT, establishing a link between RFT and local material behavior.  Our theory reproduces experimental RFT data without any parameter fitting and  generates RFT's  key simplifying assumption: a geometry-independent local force formula.  Analysis of the system explains why RFT works better in grains than in viscous fluids, and leads to an analytical criterion to predict RFT's in other materials.

\end{sciabstract}

\pagebreak
\textbf{One Sentence Summary:}
We have discovered a fundamental link between plastic flow rheology and the as-yet empirical Resistive Force Theory for immersed intruder motion.\\


\section*{Introduction}

The interaction of solid objects with a surrounding, plastically-deforming media is a common aspect of many natural and man-made processes. In the animal world, when organisms undulate, pulse, crawl, burrow, walk, or run on loose terrain they implicitly deform their environment to produce propulsive reaction forces giving rise to their motion\cite{dickinson2000animals}. The physics of such interactions have been the subject of many studies, from aquatic organisms\cite{vogel1996life,lauder2002experimental} to small insects and lizards\cite{wang2005dissecting,li2013terra} to humans and other legged-mammals\cite{thorpe2007origin, biewener1990biomechanics}. Similar principals are used for robotic applications to design machines that run\cite{bhushan2009biomimetics}, fly\cite{ma2013controlled}, swim\cite{williams2014self}, or walk in fluids or sand\cite{Ijspeert09032007,maladen2009undulatory}. Such complex interactions are also key to terramechanics of vehicular locomotion on granular substrates, models of excavation in sand and soil\cite{bekker1960off,meirion2010empirical}, and the study of similar problems in extraplanetary conditions\cite{johnson2010measurement,wong2012predicting}.  These topics and others, including cratering dynamics and penetration in plastic solids \cite{uehara2003low, shergold2004mechanisms}, all depend crucially on the way local material properties produce global resistive forces on arbitrary intruders.

Motivated by past observations in fluids \cite{lighthill75}, a simple yet very effective empirical tool known as Resistive Force Theory (RFT) for granular materials has been proposed to approximate the forces on intruding objects moving through granular media. Despite the fact that a fundamental derivation is missing, when coupled with the force balance equations, the theory provides a simple and predictive tool for simulating the locomotion of arbitrarily shaped moving bodies in loose terrain \cite{ding2011drag,maladen2011RSociety,li2013terra}. The simplicity of the theory and its predictive effectiveness are surprising in light of the complex, nonlinear, history-dependent, and oftentimes visibly nonlocal constitutive properties of granular media \cite{schofield1968critical,kamrin10a, kamrin12b, henann13,chen2013limit}.  RFT was initially developed to approximate the speed of swimming micro-organisms at low Reynolds numbers \cite{gray1955propulsion} by studying the thrust and drag of individually moving elements of its body. This method has been used extensively to study the hydrodynamics and mobility of cells as reviewed in \cite{lauga2009hydrodynamics}. However, ironically, experiments show that RFT actually works much better in granular media than in viscous fluids, a seemingly simpler material \cite{rodenborn13}. 


Is it a coincidence that such a simple model works in such a complex media or are there deeper roots to this observation?  Beyond theoretical concerns, understanding the origin of this very useful tool would help us determine what other materials induce a valid RFT and what geometric conditions limit the RFT approximation. Knowing the mechanics of RFT would enable the theory to be used in circumstances that are hard to empirically calibrate experimentally such as locomotion in micro-gravity or in circumstances where intruders have non-uniform or deformable surface properties.  Previous explorations in this field have mostly relied upon experimentally driven studies or discrete particle simulations using the Discrete Element Method (DEM). While useful as observational tools, our study herein exploits a continuum mechanical approach for its ability to make direct connections between a small set of meso-scale physical properties and the global physics of the flows they induce.

We have found that granular RFT is generated in its entirety from two of the most salient mechanical behaviors of granular media: a frictional yield criterion and an inability to support tension.   A simple constitutive model formed from these hypotheses was solved in numerous 2D and 3D situations (using the finite element method) and the results show rather conclusively that: (i) the continuum model quantitatively predicts existing experimental intrusion data and corresponding RFT input data, (ii) the model reproduces the local superposition rules postulated by RFT on global bodies, and (iii) by comparing to analogous viscous flow simulations,  the fundamental superposition concept is indeed much stronger in the granular model.  Then, by dimensional analysis of the continuum equations for granular media and viscous fluids, we discover why the RFT approximation is so strong in granular media and why it is less so for viscous fluids.  In so doing we identify new fundamental RFT formulas, which relate the RFT inputs to properties like granular density and the gravitational acceleration.

\section*{Resistive Force Theory}
In recent experimental studies of arbitrarily-shaped intruders moving in granular beds, it was determined that the resistive force against intruder motion is rather well represented by a simple \emph{superposition rule} \cite{maladen2011RSociety}; the intruder boundary can be decomposed into a connected collection of differential planar elements and the total resistive force is deemed equal to the sum of the resistive forces on each plane \emph{as if it were moving steadily on its own}.  For concreteness, let us consider for now the case of an arbitrarily shaped quasi-2D intruder having some thickness $D$ buried in a granular bed; we will generalize to 3D in the upcoming sections. Gravity points in the $\hat{z}$ direction, $z=0$ represents the granular surface. For any subset $S$ of the \emph{leading surface} of the intruder, RFT is defined by the claim that when the body is moving in the $xz$-plane the resistive force $(f_x,f_z)$ on $S$  is well-approximated by
 \begin{equation}\label{RFT1}
 (f_x,f_z)= \int_S \big(\alpha_x(\beta,\gamma),\ \alpha_z(\beta,\gamma)\big) \ H(z)\ z \ dS
 \end{equation}
 where $\beta$ is the orientation angle (attack angle) of the differential surface element and $\gamma$ is the angle of the velocity vector (intrusion angle) of the element, both measured from the horizontal (see \textbf{Fig. 2}).  Formally, a surface element is on the leading surface if a ray along its velocity vector does not intersect another point on the surface. The term $H(z) z$, for $H$ the Heaviside function, removes resistive force above the free surface and increases resistance linearly with depth to reflect increasing hydrostatic pressure. The key ingredient in RFT is the selection of the functions $\alpha_x$ and $\alpha_z$, which is done experimentally from force data for small intruding flat plates under various $\gamma$ and $\beta$ conditions. \\

 \section*{Continuum Modeling}  To predict the deformation and stresses within a continuous body of grains, frictional rheologies based on the Mohr-Coulomb (MC) yield criterion or Drucker-Prager (DP) criterion are commonly used models. We choose to study the DP yield criterion  in this paper as it is simpler to implement in 3D; in plane-strain the two criteria are equivalent.  We make three choices in closing the system of constitutive equations: (i) we choose a constant internal friction coefficient $\mu_c$, (ii) we enforce incompressible flow during yielding, which assumes a rapid approach to a critical-state of volume-conserving flow \cite{schofield1968critical} --- granular dilation is typically only a few percent regardless --- and (iii) we append an `opening rule' to the system that lets material open volumetrically when a material element attempts to enter a state of tension, to model the cohesionless granular disconnection response. Open states often occur temporarily in the wake of a moving intruder, before material from above collapses back down; see the Supplemental Materials (Section \textbf{S1}) for more details. An opening rule, while key to accurately representing many large flow processes in granular media, like the cases studied herein, is in fact a recently proposed modification in granular and soil plasticity \cite{dunatunga2015}, which traditionally restricts to dense, well-pressurized states. The bare-bones granular flow model we use does not account for rate-sensitivity, strain-dependent strength/porosity, fabric anisotropy during flow, or nonlocal effects based on particle size, which are all known to exist.  
 
 The strain-rate tensor is defined from the spatial velocity field, $v_i$, by $D_{ij} = (\partial v_i/\partial x_j + \partial v_j/\partial x_i)/2$.  We define the scalar (equivalent) shear-rate as $\dot{\gamma}=\sqrt{2{D}'_{ij}{D}'_{ij}}$ where $D'_{ij}=D_{ij}-\delta_{ij}D_{kk}/3$ is the strain-rate deviator.   Assuming that the Cauchy stress, $\sigma_{ij}$, is co-directional with the strain-rate, and that the DP yield criterion is satisfied during yielding, we write

 \begin{equation}
 \sigma_{ij} = -P \delta_{ij} + 2\mu_c P D'_{ij} / \dot{\gamma} \ \ \ \ \textnormal{if} \ \ \ \ \dot{\gamma},P>0.
 \end{equation}
 In the above, $P=-\sigma_{kk}/3$ is the isotropic pressure.  Whenever $\dot{\gamma},P>0$ we assert incompressibie plastic flow ($D_{kk}=0)$ such that the density of the packing remains at a chosen `dense' value $\rho_c$.   Whenever $\rho<\rho_c$, we set $\sigma_{ij}=0$ to represent granular disconnection. This condition is extremely important to assure granular material cannot sustain tension below critical density, which is a key difference between this approach and standard soil mechanics.  Momentum balance,  $\partial\sigma_{ij}/\partial x_{j}+\rho g_i=\rho  \dot{v}_i$,  closes the system for arbitrary boundary value problems, where $g_i$ is the acceleration of gravity and the superscript dot represents the Lagrangian time derivative.  We shall refer to this continuum model as ``frictional plasticity."  The material is represented with only two constants, $\mu_c$ and $\rho_c$.

To provide stresses in rigid zones, where $\dot{\gamma}=0$ and $P>0$, and to aid in implementing the above pressure and flow constraints numerically, we append a small elastic strain component to the deformation, which is a common method to approach the rigid plastic limit \cite{hill1956new}.  See the Supplemental Materials (section \textbf{S1}) for more details.  As long as the elastic stiffness is sufficiently high the observed plastic flow behavior is unaffected, a point we verified directly in our simulations.

 \section*{Results}
 We numerically implemented the aforementioned model in 3D using Abaqus/Explicit \cite{Abaqus}.  We first consider problems with plane-strain symmetry. No-penetration conditions are applied at the sides and bottom of the bed and the free surface is pressure-free. Gravity is first gradually applied to obtain the proper depth-dependent pressure distribution in the material before intruder motion begins. The intruding object is represented as a fully rough, thin object.  Sample flow directions and velocity distributions within the granular material obtained by theory are shown in \textbf{Fig. 1} and compared to DEM results in the literature for the same geometry having similar density and internal friction \cite{ding2011drag} . It is worthwhile to note that even though our intruder boundary condition assumes a fully rough interaction, which is an inexact representation of the condition assumed in the DEM simulation, the positive comparison suggests this difference is not crucial. 
 
 \begin{figure}
  	
  	\begin{tabular}{c c c}
  		& \multicolumn{2}{c}{\includegraphics[width=0.6\textwidth]{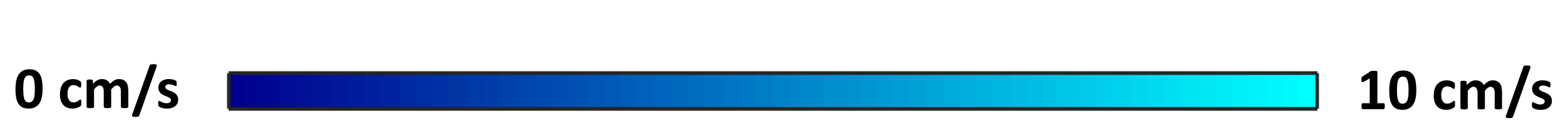}}   \\ 
  		\raisebox{12mm}{\rotatebox{90}{\textbf{Theory}}} &\includegraphics[trim= 1.6in 1.5in 1.4in .in, clip=true,width=0.42\textwidth]{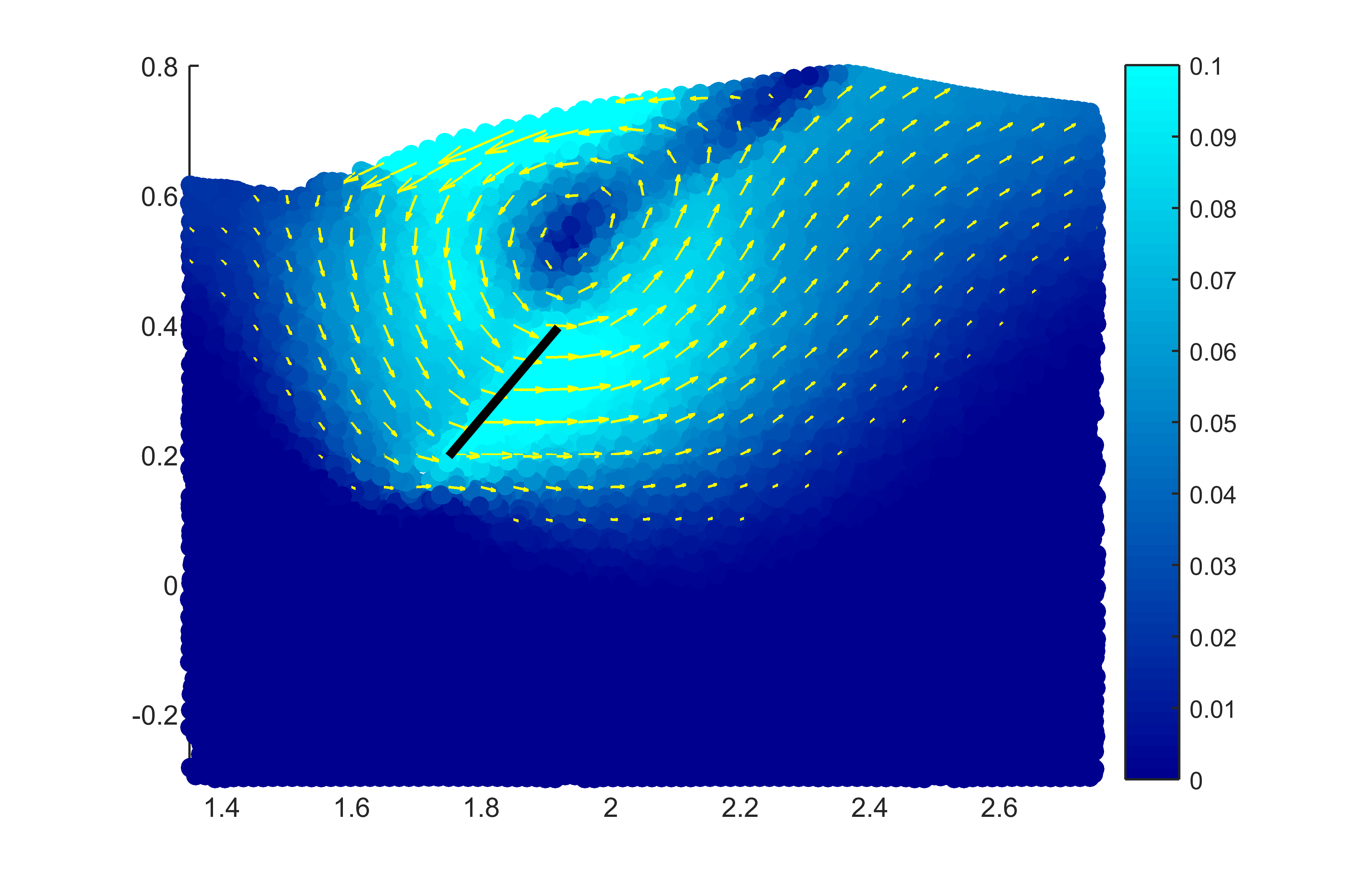}  & \includegraphics[trim= 1.65in 1in 1.35in .4in, clip=true,width=0.42\textwidth]{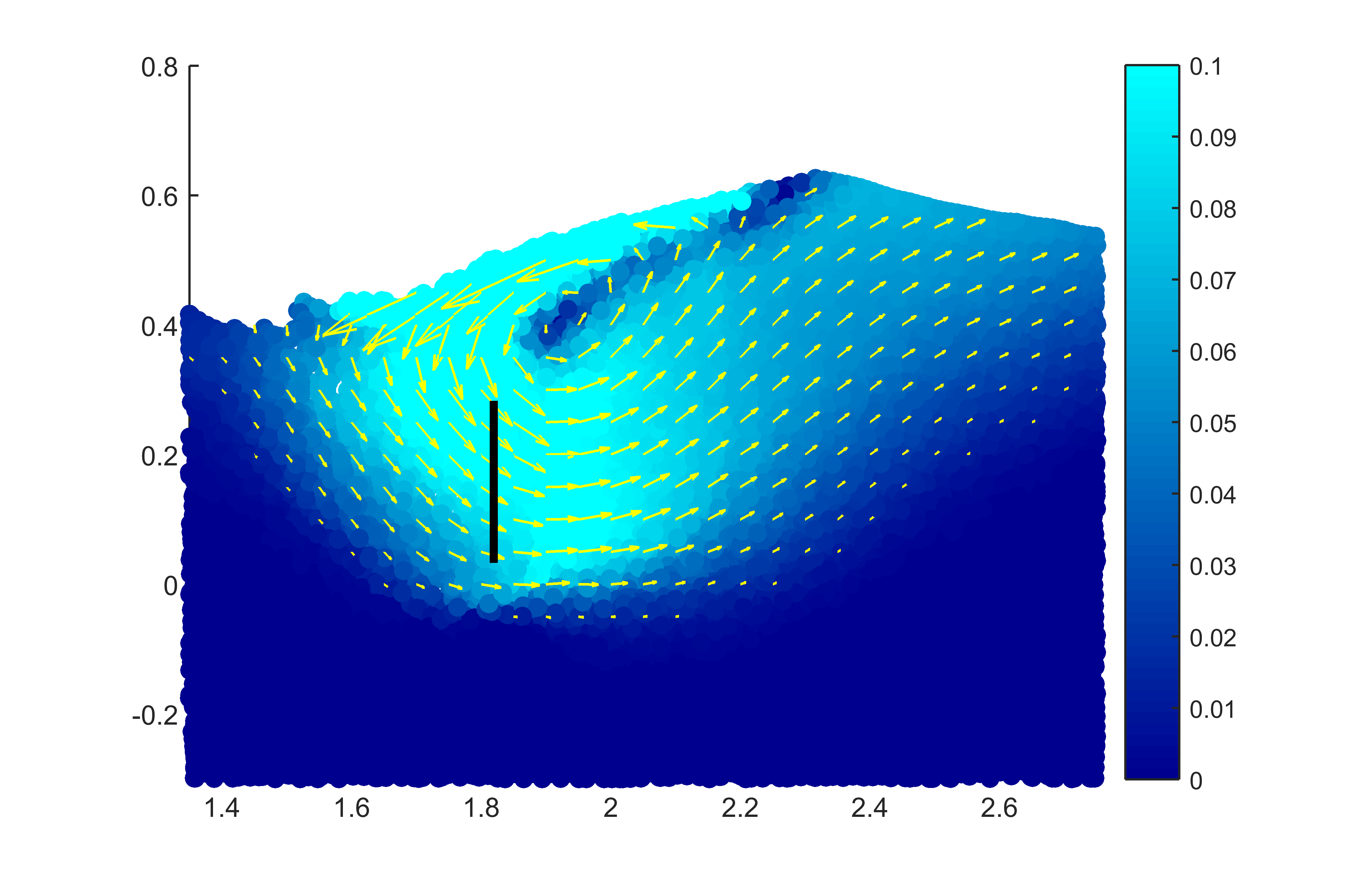} \\ 
  		
  		\raisebox{13mm}{\rotatebox{90}{\textbf{DEM}}} & \includegraphics[trim= 1.5in 1in 1.5in .in, clip=true,width=0.42\textwidth]{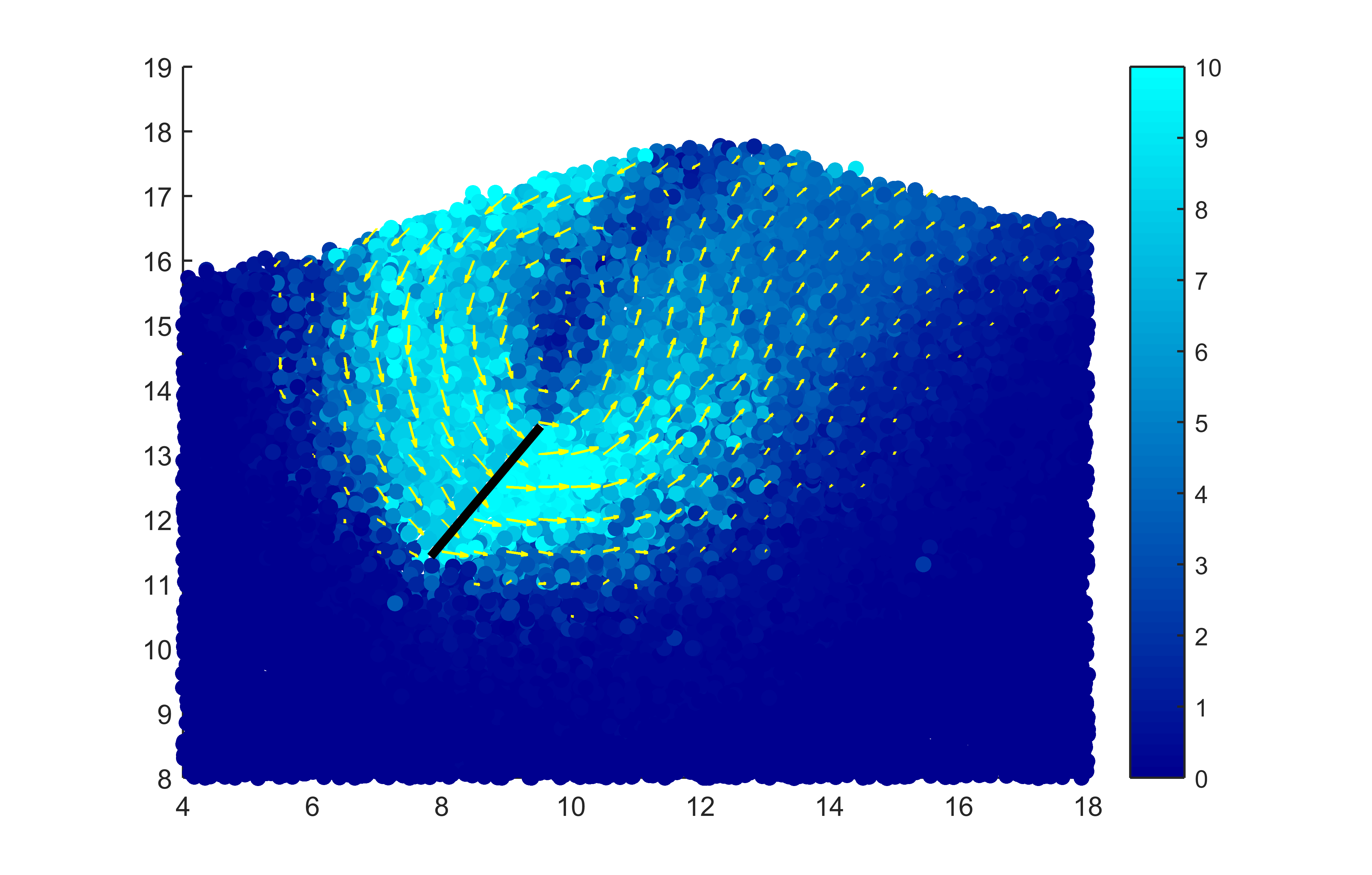} & \includegraphics[trim= 1.5in .9in 1.5in .1in, clip=true,width=0.42\textwidth]{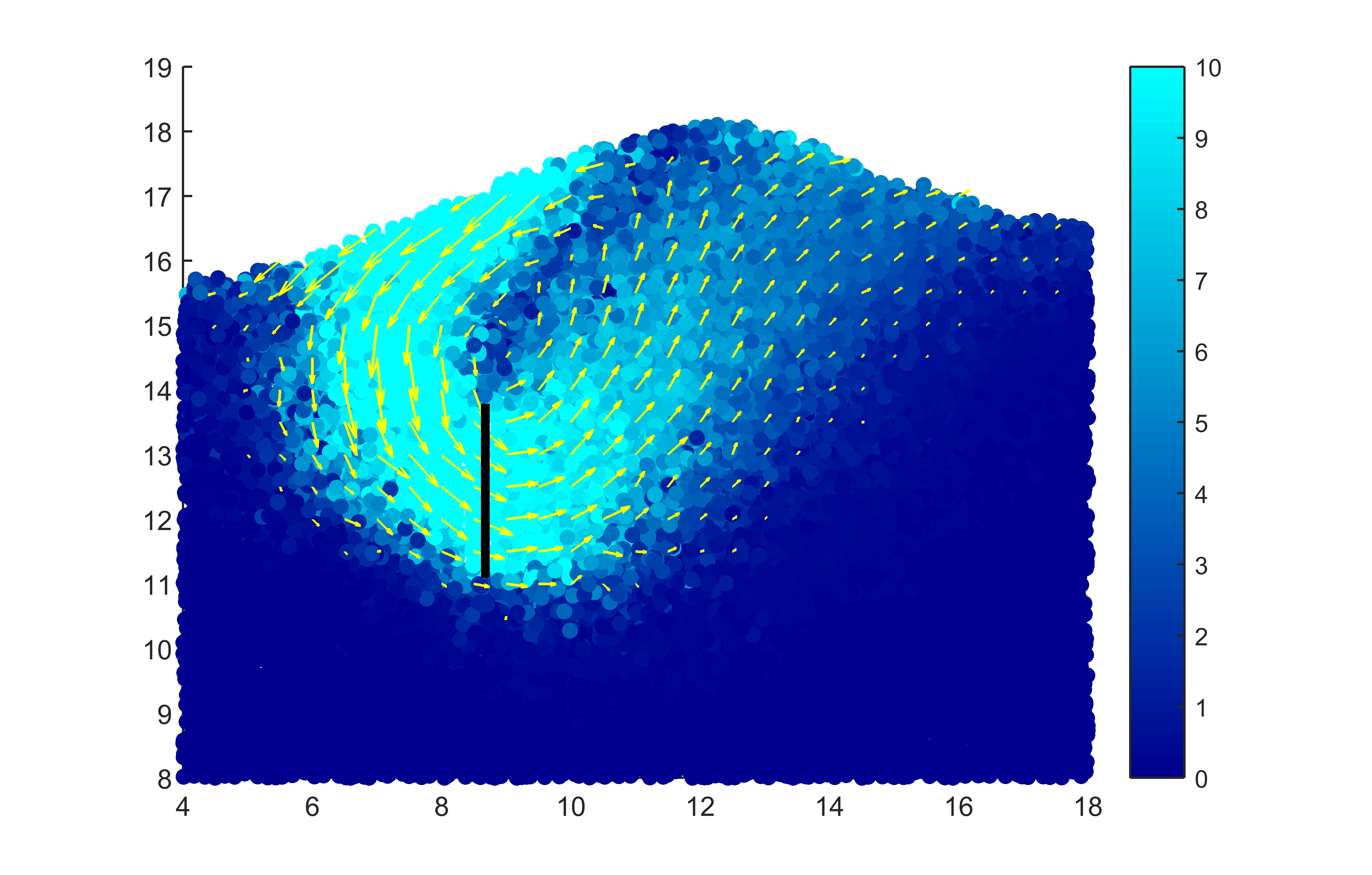} \vspace{.3in}\\
  	\end{tabular}

  	\caption{\textbf{Theoretically predicted intrusion flow fields.} Speed distribution (contours) and velocity directions (arrows) due to the  motion of a buried flat intruder of different orientations moving rightward at 10 cm/s. Results of the continuum theory (top row) and DEM simulations from the literature\cite{ding2011drag}. (bottom row)}
\end{figure}

 \begin{figure}
 	
 	\centering
 	\includegraphics[trim= 1.2in 7.8in 4.4in 1in, clip=true, width=3in,]{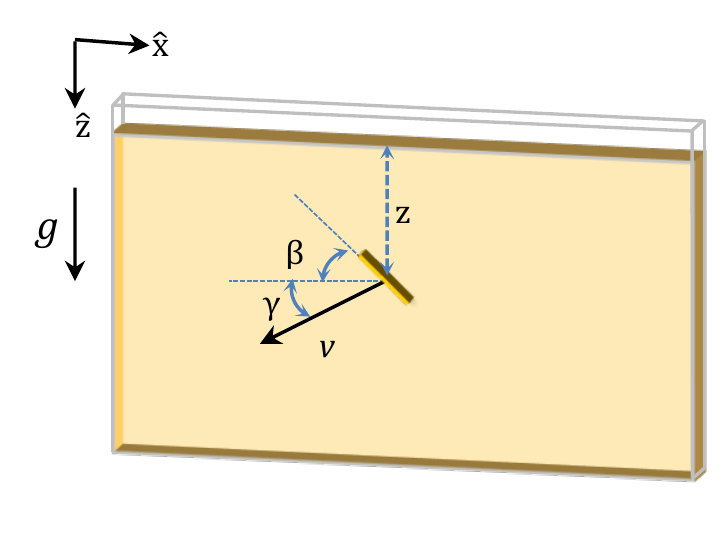}\\ \vspace{.5cm}
 	\includegraphics[trim= 1.3in 0in 1.37in .1in, clip=true, height=2in]{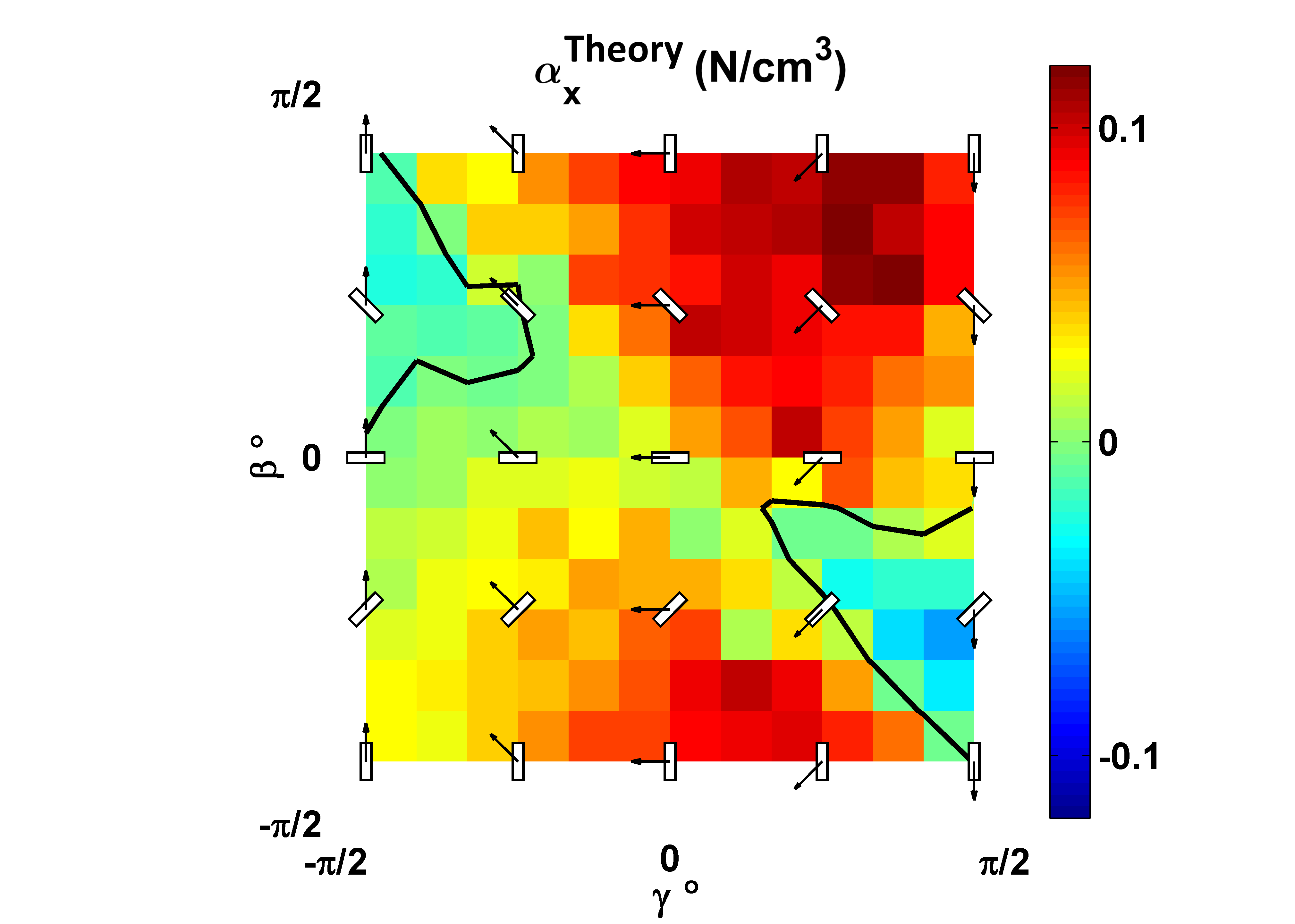} \includegraphics[trim= 1.1in 0in .4in .1in, clip=true,height=2in]{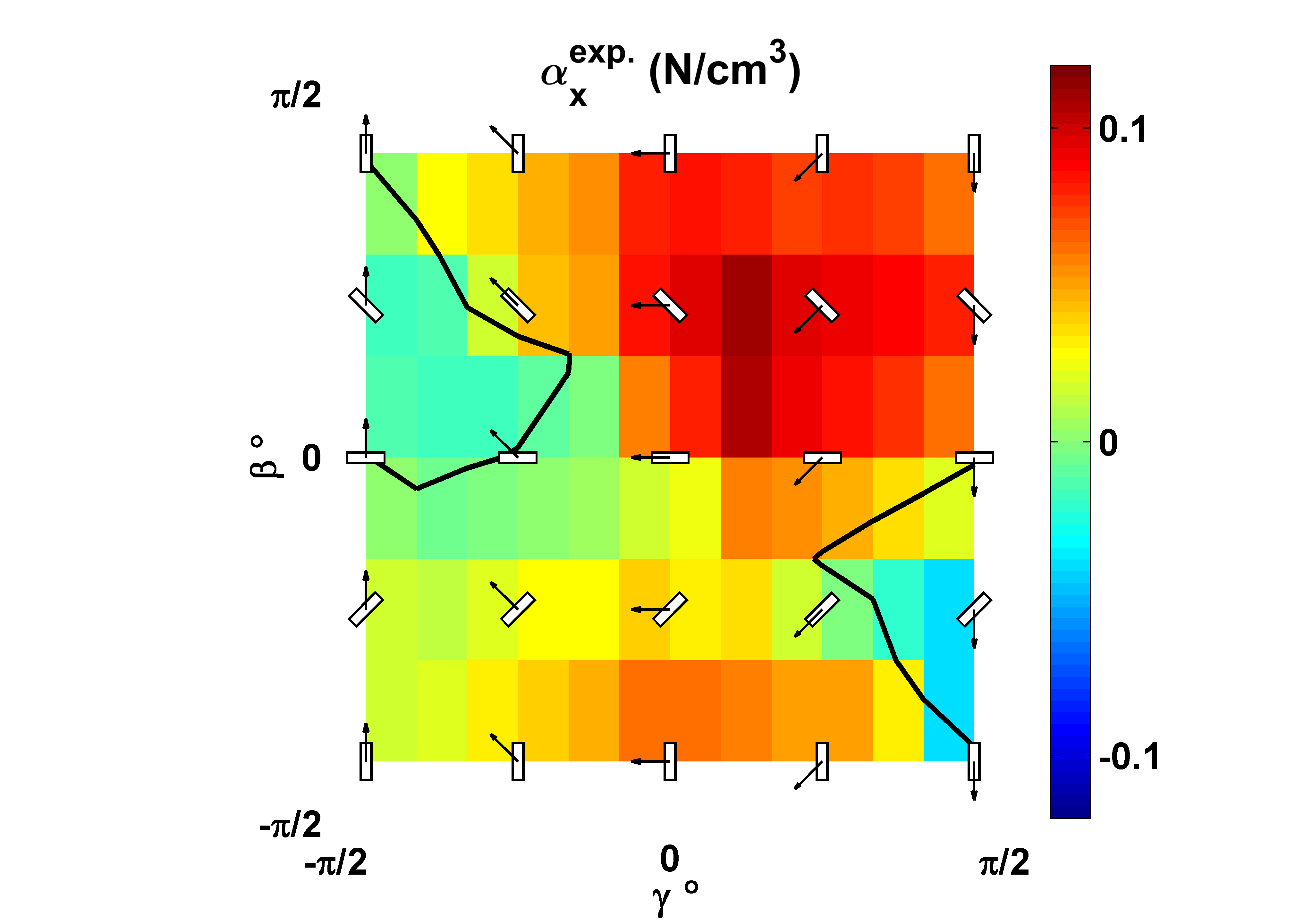}\\
 	\includegraphics[trim= 1.3in 0in 1.37in .1in, clip=true, height=2in]{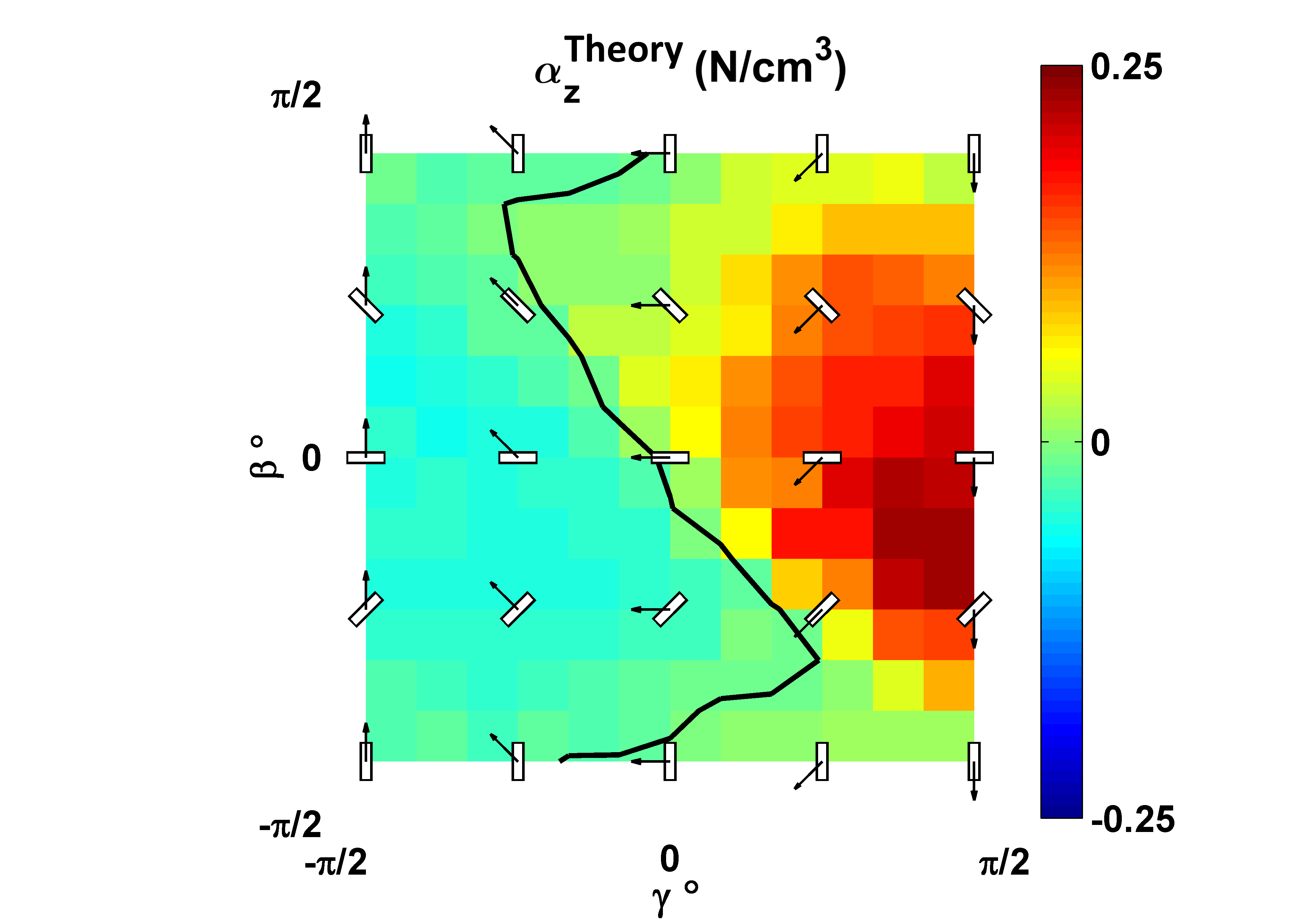}\includegraphics[trim= 1in 0in .5in .1in, clip=true, height=2in]{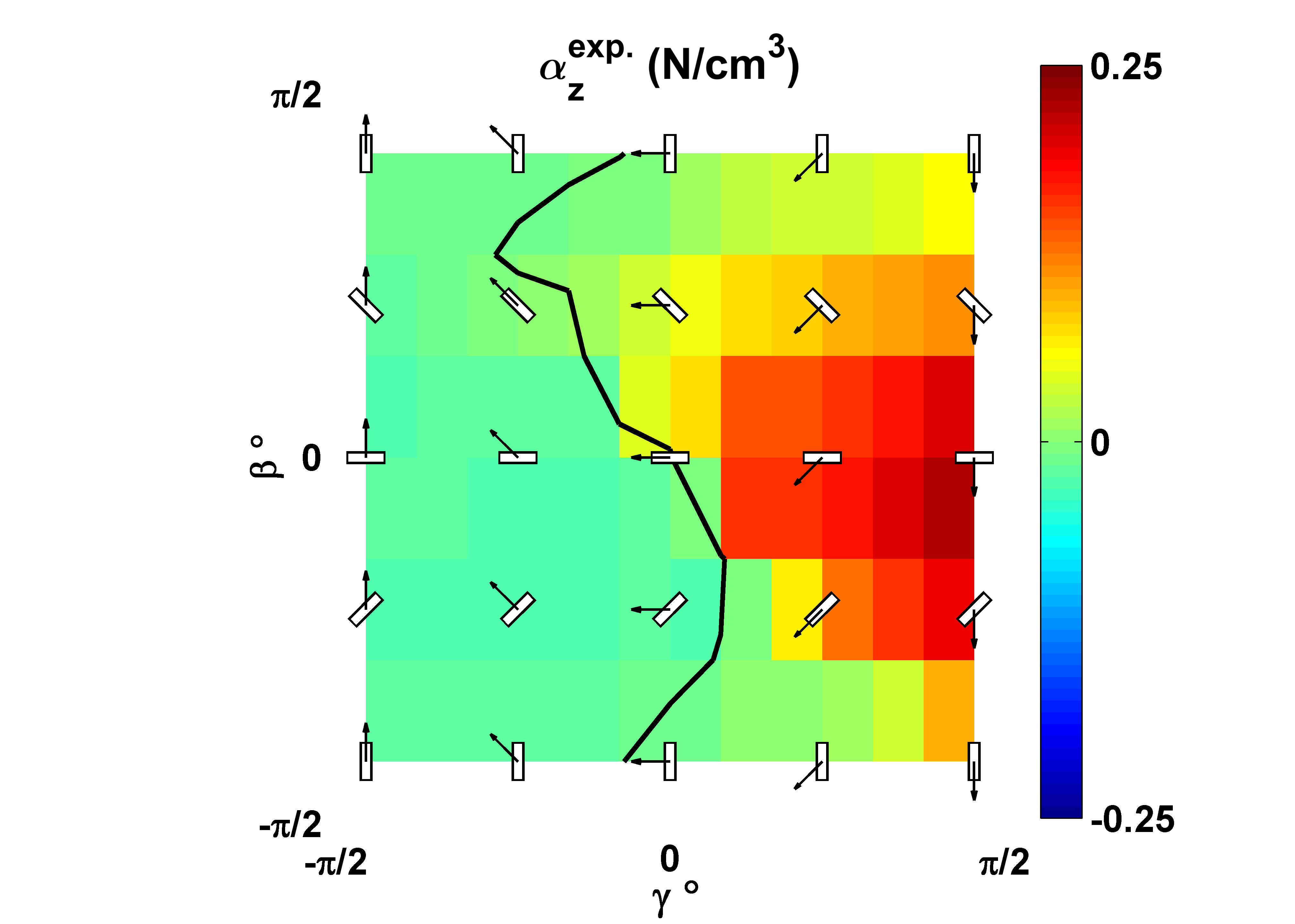}\\

 	\caption{\textbf{Theoretically predicted vs experimentally obtained resistive force plots.} RFP's obtained from frictional plasticity simulations (`Theory' superscript), compared against published\cite{li2013terra} experimental RFP's  (`exp.' superscript) for a media composed of glass beads.  Solid black lines show the zero contours.}
 	
 \end{figure}

To perform a quantitative analysis of the approach and to establish the connection between frictional plasticity and RFT, we apply the finite-element model above to a flat intruder moving under many attack angles $\beta$ and intrusion angles $ \gamma$ ($ 0 < \beta , \gamma < \pi $) to obtain predictions for the resistive force plots (RFP's), $\alpha_x$ and $\alpha_z$.  For each angle pair, the drag and lift forces acting on the plate are extracted by averaging over a period of time when plastic flow is well-developed.  The combined data set is used to populate the $\alpha_x$ and $\alpha_z$ plots.  Our model uses the properties of loose packed 3mm glass beads as reported in \cite{li2013terra}; the friction coefficient is $ \mu_c=0.38 $ based on their repose angle calculation, and $\rho_c=1.638$ g/cm$^3$, attained by assuming their solid-grain density, $\rho_s=2.6$ g/cm$^{3}$, at the close-packed packing fraction of $63\%$.  As plotted in \textbf{Fig. 2}, RFP's of $\alpha_x$ and $\alpha_z$ are strikingly similar to the experimentally obtained RFP's reported in \cite{li2013terra} for loose-packed 3mm glass beads. Furthermore the zero contour lines in the numerical RFP's follow the same path as that of the experimental RFP's. The only noticeable difference between the two sets of figures is observed in the location of maximum drag force in the $\alpha_x^{\textnormal{theory}}$ plot, which can be attributed to the no-slip intruder boundary condition used in the continuum simulations. What is more remarkable about this figure is that no fitting parameters are used in the constitutive model, only the physically measured quantities of solid grain density, repose angle, and packing fraction of the experimentally used material.

These plots also reproduce the resistive force distribution of arbitrarily shaped objects moving within the continuum granular material.  For instance, in \textbf{Fig. 3}, the force distribution and resultant forces acting on circular- and diamond-shaped intruders are obtained directly by theory and then compared with prediction from the same theory-generated RFP's shown in \textbf{Fig. 2}. Assuming the superposition rule of \textbf{Eq.1}, the resultant force is calculated and plotted against the direct-theory resultant force. These illustrations show the extent of the validity of the superposition rule of RFT and its origin within frictional plasticity. Though some errors are observed at the edges, the force distributions from both methods show a good match, while resultant force vectors show a near perfect correlation. This observation suggests that the deviations in force distributions may be due to numerical variations in the explicit FEM implementation of the theory.

 \begin{figure}
 	\centering
 	\includegraphics[trim= 1.3in 6in 4in 3in, clip=true,width=2.2in]{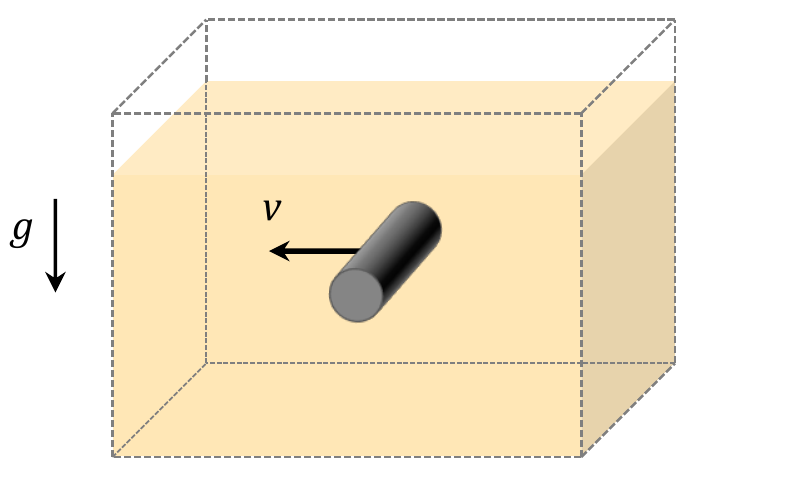}\includegraphics[trim= 2in 1.7in 2in 3.3in, clip=true,width=2in]{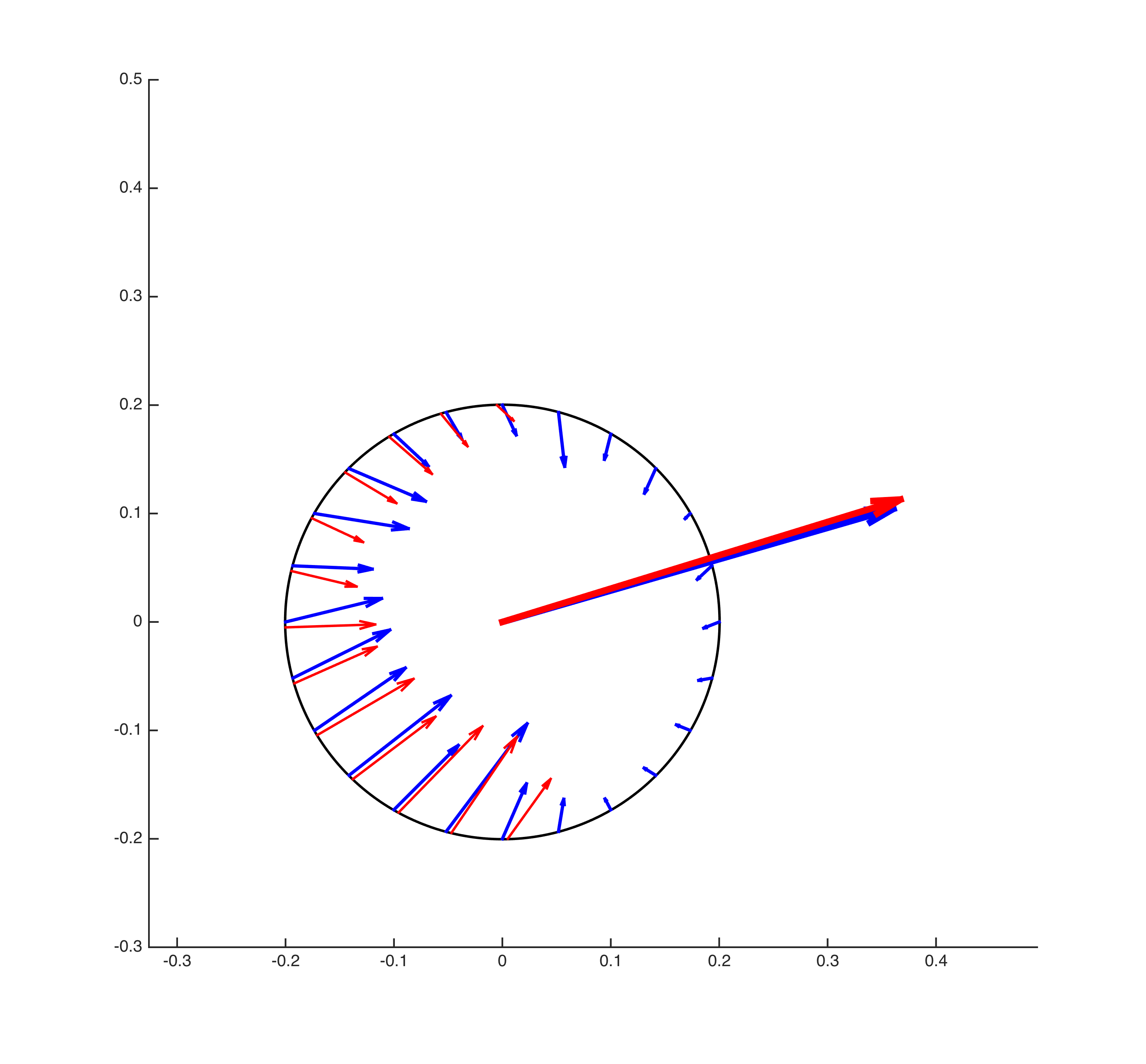}
 	\includegraphics[trim= 1.3in 8in 4in .9in, clip=true,width=2.2in]{2dillustration}\includegraphics[trim= 2.4in 2.2in 1.6in 1.6in, clip=true,width=2in]{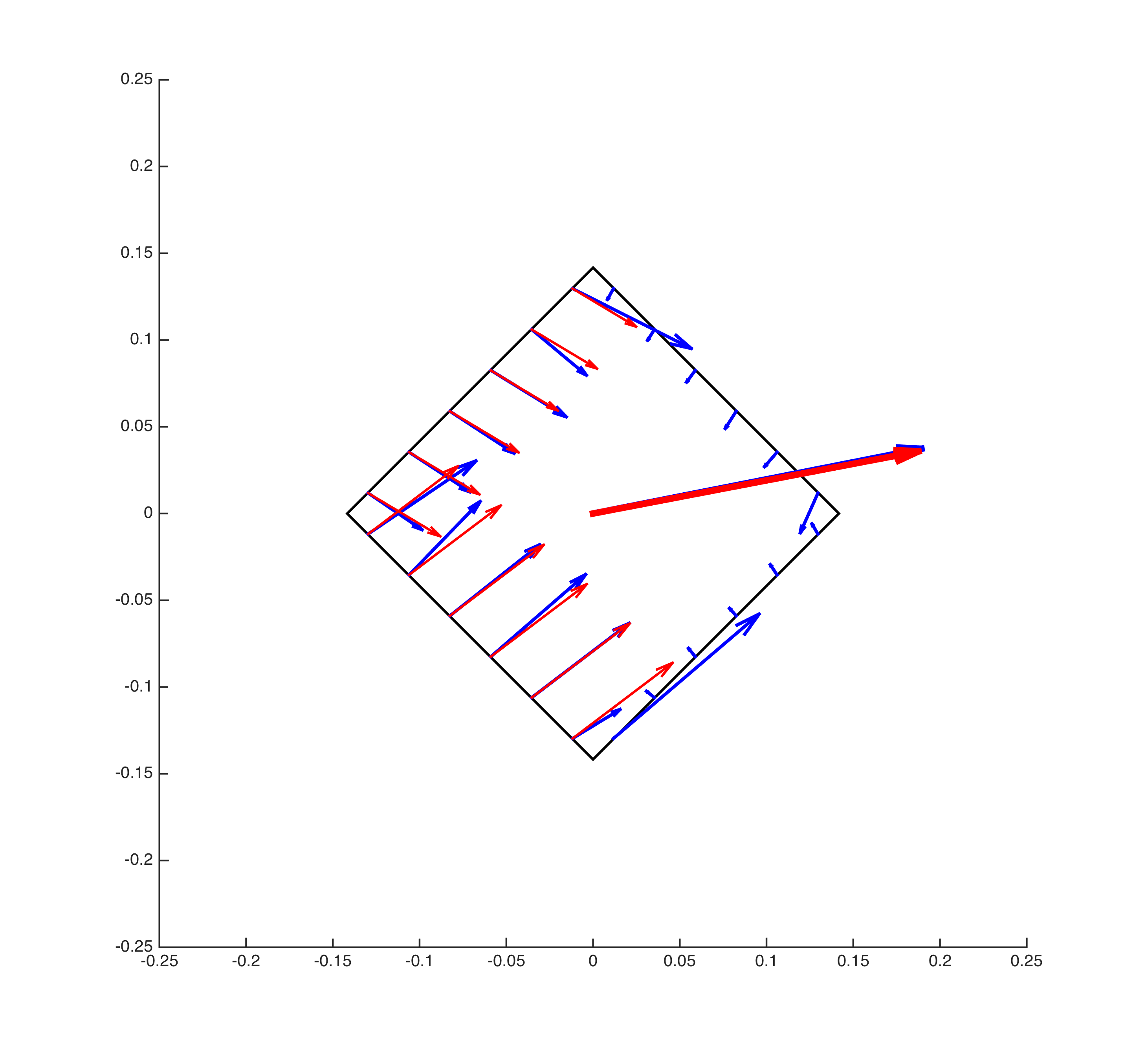}
 	
 	\caption{\textbf{Demonstration of the superposition rule arising from the flow model.} Distribution of force on the perimeter of two long moving objects as calculated directly from continuum simulations of our theory (\color{blue}$\to$\color{black}), compared to predictions from RFT using theoretical RFP's from figure (2) (\color{red}$\to$\color{black}). Net resistive force shown at the center of the object.}
 	
 \end{figure}

\textbf{Eq.1} can be extended naturally to 3D to include a non-trivial out-of-plane dimension; experiments\cite{li2013terra} have verified that surfaces whose shapes vary in the $y$ direction also maintain the superposition principle. To assess the generality of the RFT superposition principle in 3D, we employ our model to study forces on a sequence of buried V-shaped intruders.  Because RFT is supposed to apply only to the leading surface of an intruder,  we limit our attention to an obtuse V geometry, with angle fixed at $\theta_V = 135^\circ$, in which most orientations of the V admit both wings to be part of the leading surface.

 \begin{figure}
 	\hspace{1.8in} Frictional plasticity \hspace{1.2in} Viscous fluid \\
 	\includegraphics[trim= 0in 0in 0in 0in, clip=true, width=6.2in]{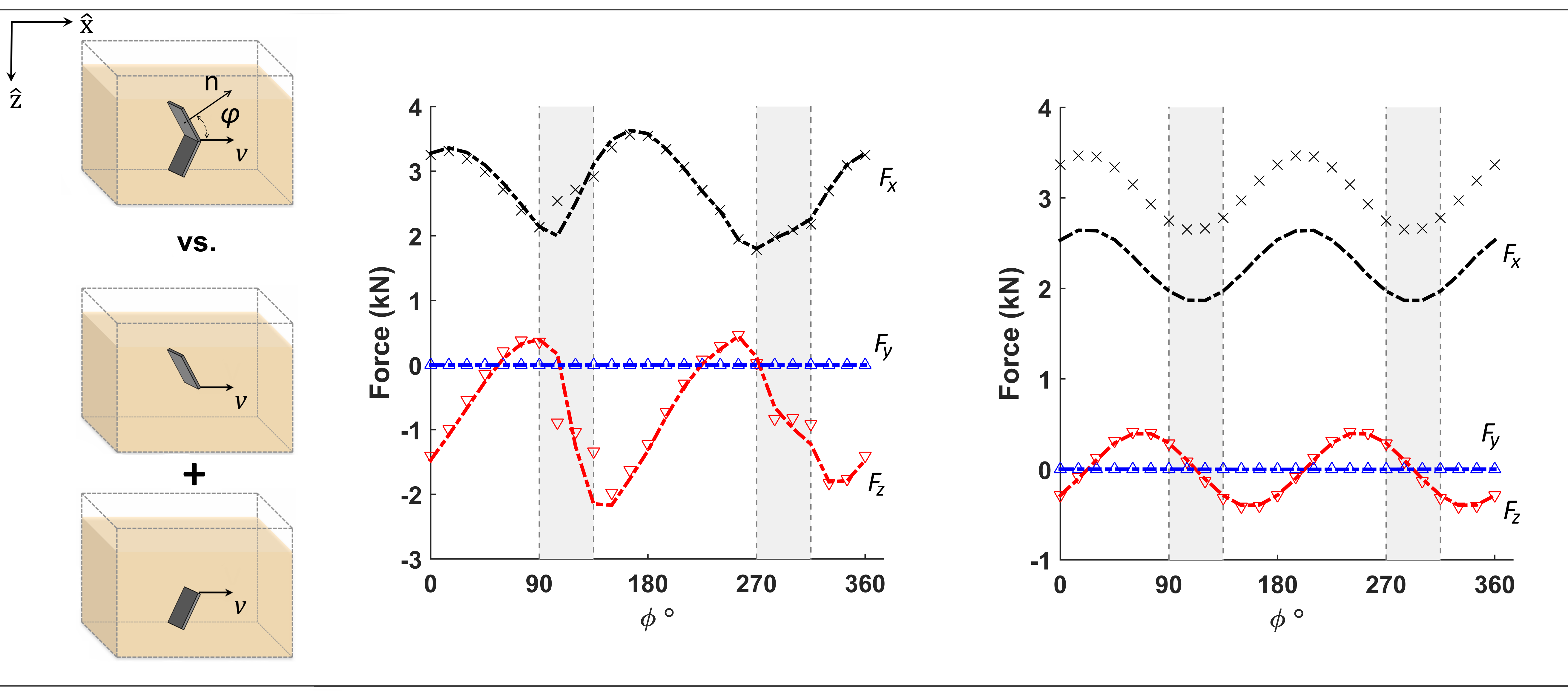}
 	\includegraphics[trim= 0in 0in 0in 0in, clip=true, width=6.2in]{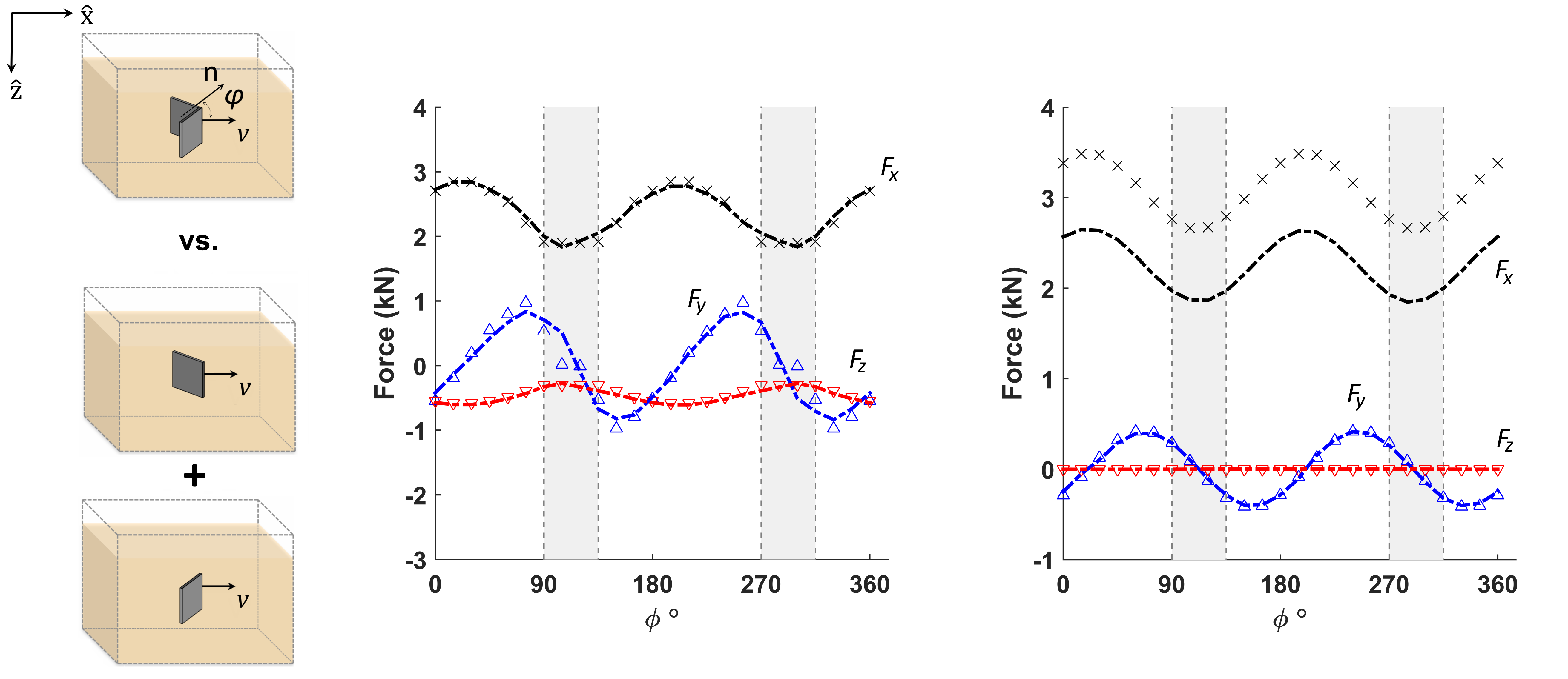}

 \caption{\textbf{Validity check for superposition in 3D.} The drag force, $F_x$, acting on a rightward moving submerged  `V' intruder  ($\times$),  and the components of lift, $F_z$ (\color{red} $\bigtriangledown$\color{black}), and lateral force, $F_y$ (\color{blue} $\triangle$\color{black}), compared with superposition values (dashed) at various orientation angles, $\varphi$, for vertical (top row) and horizontal (bottom row) intruder alignments in granular flow (left, $\rho=4 $g/cm$^3$, $\mu=$0.4, depth=0.6m) and viscous fluid (right, $\eta v$=15N/m). Intruder consists of two square plates of side length 20cm.  Grey regions indicate orientations where the plates are not both on the leading surface of the V.}
 
  \end{figure}
 
We consider the case where the intruder is moving as a one piece object and we record the total resistive force as a three-dimensional vector. Then we run two new simulations, one for each wing of the intruder, where the wing is simulated as a single plate maintaining its same exact alignment, positioning, and motion that it had when it partook in the group V motion. We consider both vertically and horizontally aligned intruder cases, varying the orientation angle $\varphi$ denoting a pitching angle (in vertical case) or yawing angle (in horizontal case). For the sake of comparison, we also perform the same sequence of tests assuming the surrounding media is a Stokesian fluid (see \textbf{Fig. 4}) using a similar finite-element implementation. All cases engage considerable drag force in the direction of motion, as expected. The vertically aligned intruders experience vanishing $F_y$ due to symmetry, and $F_z$ switches sign at certain orientations due to the plowing nature of the intruder movement, similar to the oscillatory nature of the $y$-force for the horizontal intruder. In the horizontal case, unlike the zero viscous $F_z$ force, the intruder in the granular bed experiences an $F_z$ force that on average pushes upward, demonstrating the well-known drag induced lift effect\cite{ding2011drag} in granular systems.

The comparison in \textbf{Fig. 4}  shows that the theory works extremely well in granular materials for all force components.  The agreement is roughly in the same range as the deviations observed in past granular RFT experiments\cite{ding2011drag, goldman14}.  In ranges indicated in grey, the leading-surface assumption of RFT is violated; i.e. one plate is behind the other.  The agreement is not as strong in these zones but still overall good. 
For this 3D study there are no analogous experimental tests in the literature to verify the findings of the theory.  The error of the RFT force prediction, $|\mathbf{F}-\mathbf{F}_{RFT}|/|\mathbf{F}|$, is found to be 7.3\% averaged over all orientations of the V and 3.9\% when orientations that violate the leading-surface assumption are excluded.

In contrast, intrusion through a viscous fluid half-space provides a considerable offset in the drag direction, which is the largest and typically most important component. The total error of the force is about 36\% averaged over all orientations. The nontrivial force component in the lateral direction actually  seems to show a good degree of superposition but this may be a coincidence; when the V angle is varied, the disagreement in this component becomes much more pronounced. In the presumably simpler case of $\theta_V =180 ^\circ$, i.e. a flat intruder, the error for the viscous superposition has significant values in both drag and lateral directions, leading to an average error of 42\% whereas the granular superposition maintains its accuracy with an average error of about 6.5\% (see Supplemental Materials, section \textbf{S4}) We evaluate RFT in 3D by comparing the forces obtained directly from the V-shaped intruder (\textbf{Fig. 4}  solid lines) with the forces obtained by summation of the forces from the two separate, individually moving component plates (\textbf{Fig. 4}  dashed lines). This comparison shows that the theory works extremely well in granular materials for all force components. We find the agreement is more pronounced when a particular wing is positioned such that there is less interaction between the flow-fields stemming from each plate.  The agreement is roughly in the same range as the deviations observed in past granular RFT experiments \cite{ding2011drag, goldman14}. For certain orientations, the disagreement is due to a `shadowing' effect caused by the leading wing on the trailing wing. For instance at  $\varphi=0^{\circ}$  a good agreement between group and individual forces is observed for the vertically aligned intruder. One can expect the flow fields of both wings to point in the same direction (see \textbf{Fig. 1}) and little effect should be expected from the cross-interaction of the fields. But once the same object is rotated to $\phi=90^{\circ}$, these fields would interact and therefore the disagreement in forces is expected to increase in all components, as confirmed in the figure. For this 3D study there are no analogous experimental tests in the literature to verify the findings of the theory. However, the results shown in \textbf{Fig. 4} suggest that for 3D objects that do not contain sharp acute angles and whose geometry does not shadow other parts of its own perimeter, reasonable agreement should be expected if granular RFT is used. The error of the RFT force, $|\mathbf{F}-\mathbf{F}_{RFT}|/|\mathbf{F}|$, is found to be 7\% averaged over all orientations of the V.
 
In contrast, intrusion through a viscous fluid half-space provides a considerable offset in the drag direction, which is the largest and typically most important component. The total error of the force is about 36\% for all orientations, although the nontrivial force component in the lateral direction actually  seems to show a good degree of superposition.  In light of the error in the drag direction, we cannot explain this agreement nor rule out coincidence; when the V angle is varied, we find the disagreement in this component becomes much more pronounced. In the presumably simple case of $\theta_V =180 ^\circ$, i.e. a flat intruder, the error for the viscous superposition has significant values in both drag and lateral directions, leading to an average error of 42\% whereas the granular superposition maintains its accuracy with an average error of about 6.5\% (see Supplemental Materials, section \textbf{S4})

\section*{Analytical explanation}  The results thus far have demonstrated numerically that the hypotheses of RFT emerge relatively strongly from the continuum equations of frictional plasticity.  To provide more of an explanation as to \emph{why} these equations replicate RFT, we consider the following illustrative example, reminiscent of a ``garden hoe''. As delineated earlier, exact solutions to the continuum plasticity system are highly nontrivial to obtain, but in the example we highlight below the continuum result can be inferred using dimensional analysis without having to solve the differential equation system.

    	\begin{figure}
    		\includegraphics[trim= 1in 7.5in 2.4in .7in, clip=true, width=6in]{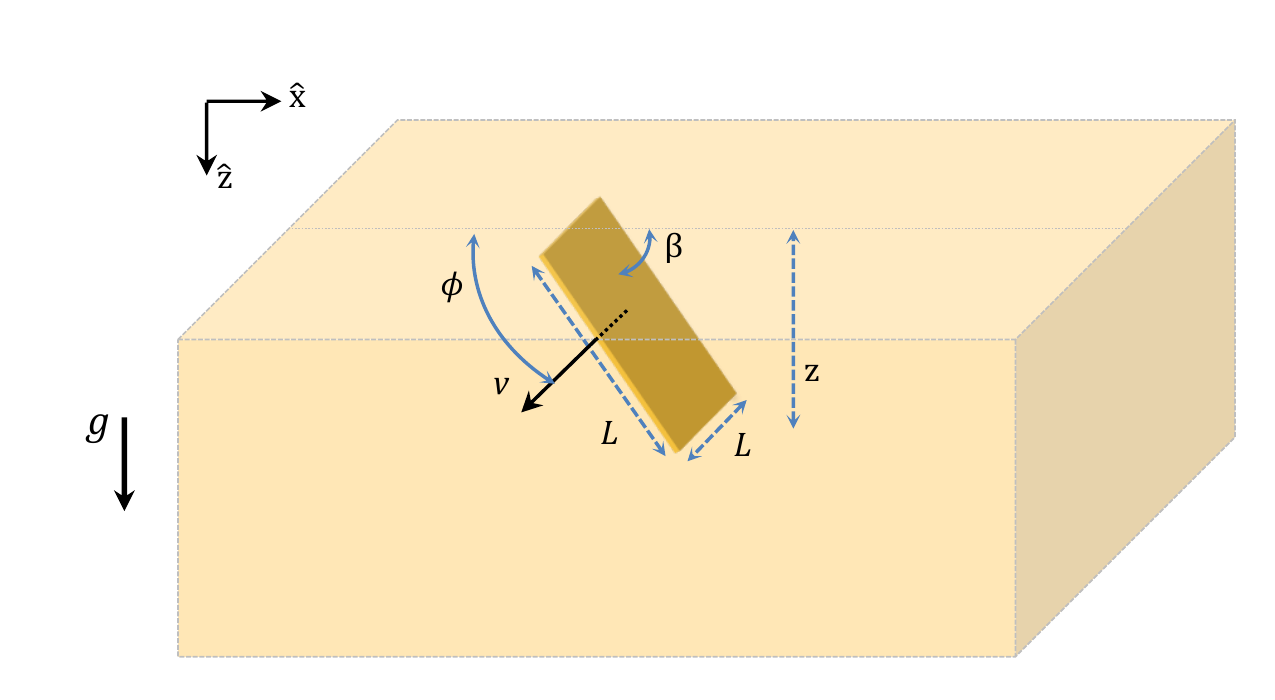}

    	\caption{\textbf{Schematic of an RFT litmus test geometry.} The `garden hoe' test geometry; an arbitrarily oriented square intruder with one edge at the free surface is set into motion in an arbitrary direction in a semi-infinite domain of material.}
    	\end{figure}    	
    	
 Consider a semi-infinite half-space of frictional continuum media (see \textbf{Fig. 5}).  Suppose a square-shaped intruder with edge length $L$ is inserted into the media at an angle $\beta$ from the horizontal. Its top edge remains coincident with the free surface. The intruder is then translated at a speed $v$ in a direction angled $\gamma$ from horizontal, producing an assumed quasi-static motion of the material.  At the moment the motion begins, the total resistive force $\mathbf{F}_{\text{gran}}$ on the intruder can be calculated from the continuum model. Due to the parameters of the model, the resulting force can only depend on  $\beta$, $\gamma$, $L$, the granular weight density, $\rho_c g$, the internal friction,  $\mu_c$, and perhaps a wall friction coefficient  between the intruder and the material, $\mu_w$.  Define $L$, $L/v$ and $\rho_c g *L*(L/v)^{2}$ to be the units of length, time, and mass, respectively.  Non-dimensionalizing, we find the dimensionless groups in the problem are $\beta$, $\gamma$, $\mu_c$, $\mu_w$ and the dimensionless force $\mathbf{F}_{\text{gran}}/\rho_c g L^3$.  We infer that the dimensionless force must equal a vector-valued function of the dimensionless inputs, implying that $\mathbf{F}_{\text{gran}}=\rho_c g L^3\ \Psib_{\text{gran}}(\beta,\gamma,\mu_c,\mu_w)$. Redefining $\mathbf{\Psi}_{\text{gran}}$ by absorbing the material constants, we have
 \begin{equation}
 \mathbf{F}_{\text{gran}} = L^3\ \mathbf{\Psi}_{\text{gran}}(\beta,\gamma).
 \end{equation} 
To use RFT one first computes RFP's of $\alpha_x$ and $\alpha_z$ by gathering data on the intrusion of a small surface element of edge length $\lambda\ll L$ as it displaces in different directions and with different orientations. For a frictional-plastic material, the above formula implies the answer must be $\big{(}\alpha_x(\beta,\gamma),\alpha_z(\beta,\gamma)\big{)} =2\mathbf{\Psi}_{\text{gran}}(\beta,\gamma)/\sin\beta$.  The RFT superposition principle (\textbf{Eq.1}) can then be used to estimate the drag force on a garden hoe of size $L$ by superposing the local drag rule over the surface.  We find
 \begin{equation}
 \mathbf{F}^{\text{RFT}}_{\text{gran}}=\int_S  \big{(}\alpha_x(\beta,\gamma),\alpha_z(\beta,\gamma)\big{)} \ z \ dS = L^3 \ \mathbf{\Psi}_{\text{gran}}(\beta,\gamma) .
 \end{equation}
This formula is precisely that of \textbf{Eq.3}. We conclude that RFT predictions exactly match those of full solutions of the frictional-plastic continuum model in the test case of the garden hoe family of geometries. Additionally, the effect of parameters such as gravity and density can be lumped as a scaling constant in representation of the vector function $\mathbf{\Psi}_{\text{gran}}$. Likewise, we predict that the resistive force plots must scale linearly in $\rho_c g$ for fixed friction constant(s).  We validated this hypothesis in additional finite-element simulations.
 
For comparison, consider a viscous, incompressible fluid rather than a granular material, which obeys the Stokes equations, $\eta\ \partial^2v_i/\partial x_j^2- \partial P/\partial x_i+\rho g_i=0$, with incompressibility condition, $ \partial v_i/ \partial x_i=0$, and  dynamic viscosity $\eta$. For the same intruder  and motion described above,  the intrusion force, $\mathbf{F}_{\text{visc}}$, now must depend on $ \beta, \gamma, v, L$, and $\eta$. It cannot depend on $\rho g$ as this term can be removed by absorbing it into the pressure in the Stokes equations without altering the resultant intruder force. We choose $L, L/v,$ and $ \eta L^2/v^2$ as the length, time and mass units respectively, giving the dimensionless variables $\beta, \gamma$, and $\mathbf{F}_{\text{visc}}/\eta L v$.  Consequently, in viscous media, the force on the intruder must have the form $\mathbf{F}_{\text{visc}}= \eta L v \ \Psib_{\text{visc}} (\beta, \gamma)$
for some vector-valued function $\Psib_{\text{visc}}$.  Absorbing the viscosity into $\Psib_{\text{visc}}$, we have
\begin{equation}
\mathbf{F}_{\text{visc}}= L\,  v \ \Psib_{\text{visc}} (\beta, \gamma).
\end{equation}
That the drag force grows linearly with $L$ is similar to other Stokesian drag formulas; the drag on a sphere is proportional to its radius, for example.
 
Similar to the previous argument for granular RFT, we can compare the above form to the one we would obtain from viscous RFT, and determine if the two can be in agreement. The resistive force theory in fluids has been historically used in describing forces on slender bodies. Within this theory, coefficients of normal and tangential drag are locally defined. In general, one computes the drag on a small characteristic length $\lambda$ at various orientations to the motion, and then estimates the force on the whole body by integration of the local drag formula over the complex path of the body.  In a higher-dimensional framework, one isolates an areal patch of small characteristic length $\lambda$ oriented at various angles $\beta$ and traveling at angles $\gamma$, and computes a local drag law, which takes the form $\{\text{Force/Area}\}= \big{(}a_x(\beta,\gamma),a_z(\beta,\gamma)\big{)} \ v$.  Unlike the granular case, viscous rheology is pressure-insensitive, which is why the local drag rule is independent of depth.  The formula for viscous drag force above implies the answer must be $\big{(}a_x(\beta,\gamma),a_z(\beta,\gamma)\big{)}= \Psib_{\text{visc}} (\beta, \gamma)/\lambda$. Similar to \textbf{Eq.1}, the total force is then estimated by superposition of the local drag rule over the surface of the macroscopic geometry:
  \begin{equation}
  \mathbf{F}^{\text{RFT}}_{\text{visc}}=   \int_S  \big{(}a_x(\beta,\gamma),a_z(\beta,\gamma)\big{)} \ v \ dS = \frac{L^2}{\lambda}v \ \mathbf{\Psi}_{\text{visc}}(\beta,\gamma) .
  \end{equation}

 We observe that the force on a plate moving in viscous fluid scales as $L^2$ under RFT assumptions, but this is not correct; per \textbf{Eq.5}, it should scale as $L$. This difference manifests due to a non-removable factor of $\lambda$, the selected micro-size, showing up in the local drag rule.  This issue causes similar errors in the common application of RFT to slender bodies such as flagella and long micro-organisms.  Because the local drag coefficients cannot be divorced from a chosen characteristic length, they are only an approximation to a local rule, and hence do not admit a strong superposition. In the Supplemental Materials (section \textbf{S2}), we explain how the accuracy of granular RFT continues even in the limit of a slender body, while viscous slender body RFT, by contrast, retains this small-scale geometrical dependence.

The above demonstrations suggest the ability of the garden-hoe test to determine which material rheologies have the potential to admit a strong RFT approximation.  The test shows that granular RFT is exact in these geometries, but does not explain its general effectiveness for curved, submerged surfaces. In more general geometries, we hypothesize that the reason for continued RFT-type superposition may be linked to the fact that the equations of frictional plasticity for planar, quasi-static, rigid-plastic problems become a \emph{hyperbolic} system in space. The stress characteristics extend from boundaries along curves saturating $|\tau|/P\le\mu_c$, that form the `slip-lines' \cite{nedderman}.  This produces domains of dependence in the material in a sense that stresses in certain zones can be attributed to the traction on a specific part of the surface of the intruder. This is in sharp contrast to viscous fluids, for example, in which the equations are \emph{elliptic}, and the stress at any point on an object's surface is influenced by the motion and shape of the entire boundary of the object.  

\section*{Discussion}
The granular constitutive model we have used in this work was chosen to capture the salient granular flow behaviors --- a frictional, cohesionless constitutive relation --- and we have shown that this bare description is sufficient to bring about RFT.  Evidently there is more to the rheology of granular flow than these essential behaviors and there are certainly limits where additional effects can become significant (see Supplemental Materials, section \textbf{S3}, for more discussion).  For example, rate dependence can become significant when the inertial number, $I = \dot{\gamma} d \sqrt{\rho_s/P}$ is sufficiently large (i.e. $I\gtrsim0.1$) \cite{dacruz05}.  Size-effects can become important in the limit where the grain size $d$ and length of the object $L$ are comparable \cite{kamrin12b}. The transients in the evolution of $\mu_c$ during flow can matter if the strain induced by the intruder over a time characterizing changes in the direction and elevation of the intruder is smaller than the critical-state strain $\varepsilon_c$; this is most noticeable in the weakening of the RFT approximation during motion reversal \cite{ding2012mechanics}.  There is no straightforward way to discern the variation in the RFP's when these effects become more pronounced, nor is it clear that RFT would be valid at all, theoretically or experimentally, in these limits.  However, because the two essential properties used in our approach are still the foundations of these more detailed constitutive approaches,  any additional effects from these sources, for intrusion problems in the typical regimes discussed in this paper, should be within the margin of error between our theoretically obtained RFPs and the experimentally reported values in the literature.

By rooting RFT in continuum mechanics, its underlying physical basis is substantiated, and at the same time we have revealed a previously unidentified collapse within the equations of plasticity.  In particular, we have shown that the surface superposition rule and the other hypotheses of RFT arise as a consequence of a local frictional yield criterion augmented with an opening ability, a point directly verified in plane-strain and full 3D numerical simulations.  Moreover, the plasticity system, when fit to the repose angle of the physical media of \cite{li2013terra}, quantitatively replicates  the empirical input data that was previously used for RFT, and the system replicates the force superposition concept foundational to RFT.  We have identified how the exact form of RFT arises in a model problem, based solely on dimensional analysis.  In so doing, we have also explained why RFT works better in granular matter than in viscous fluids, using a case study of a V-shaped intruder as well as dimensional and superposition arguments.  This new fundamental understanding of the physics underlying RFT paves the path forward to better understand and derive RFT's for other flowable materials, and apply the method in systems that might be difficult to model based on direct experimental inputs.\\


\textbf{Acknowledgements} K. Kamrin and H. Askari gratefully acknowledge support from Army Research Office Grants W911NF-14-1-0205 and W911NF-15-1-0196.

\pagebreak
\begin{center}
\section*{Supplementary Materials}
\end{center}
\vspace{1cm}
 The goal of these supplemental materials is to provide details about the modeling techniques and calculations presented in the main manuscript of this paper.  We describe different components of the finite element approach  followed by a more detailed discussion of the dimensional analysis and further numerical demonstrations.
 \vspace{1cm}
 \section*{S1. Numerical solution procedure}
 \subsection*{Elasticity-augmented constitutive relation}
 As outlined in the main paper, we describe the deformation behavior of the granular material by using a non-hardening Drucker-Prager (DP) yield criterion (constant $\mu_c$).  When implementing the model, we assume a small portion of the deformation is elastic, which closes the system mathematically in regions of non-flowing material, and provides a natural route to implementing the pressure constraints described.  The model then takes the form of a simple hypo-plastic formulation. At any time, the deformation gradient $F_{ij}$ is used to construct the velocity gradient $\partial v_i / \partial x_j   = \dot{F}_{ik} F^{-1}_{kj} $, which is divided into a symmetric stretching component $D_{ij}$ and anti-symmetric spin component $W_{ij}$. The dot represents the Lagrangian time derivative.  The stretching is further decomposed into elastic and plastic parts $D_{ij}=D_{ij}^e+D_{ij}^p$. Denoting the elastic shear and bulk moduli of the granular material by $ G$ and $K$ respectively, the constitutive relationship for the stress evolution of dense material (i.e. $\rho\ge\rho_c$) is established by a Jaumann objective rate form as follows:
 \begin{equation}
 	\begin{cases}\tag{S1}
 		\sigma_{ij}=0 ,&\text{if}\ \rho<\rho_c\\
 		\dot{\sigma}_{ij}- W_{ik} \sigma_{kj}+\sigma_{ik} W_{kj} = K D_{kk}^e\delta_{ij} +2G (D_{ij}^e-D_{kk}^e\delta_{ij}/3) ,&\text{if}\ \rho\ge \rho_c
 	\end{cases}
 	\label{objrate}
 \end{equation}

 \noindent
 where $\sigma_{ij}$ is the Cauchy stress as described in the main paper. This relationship warrants that the material is stress-free if density falls below a certain critical density $\rho_c$, indicating a disconnected or `open state' material point. Further discussion about this condition is followed in section 1.3 of this supplement.  The Cauchy stress is further divided into a hydrostatic pressure part $\bar{\sigma}=\sigma_{kk}/3=-P$ and a deviatoric part $\sigma_{ij}^{'}=\sigma_{ij}-\bar{\sigma}\delta_{ij}$, which is used to define the equivalent shear stress, $\bar{\tau}=\sqrt{\sigma_{ij}^{'}\sigma_{ij}^{'}/2}$. The plastic strain-rate, $D^p_{ij}$, is uniquely defined to ensure the following conditions: (1) $D^p_{ij}=\lambda \sigma'_{ij}$,  (2) $\lambda=0$ if $\bar{\tau}<\mu_c P$, (3) $\lambda\neq 0$ only if $\bar{\tau}=\mu_c P$, and (4) $\bar{\tau}\le \mu_c P$.  The next section details how this system of updates and constraints is implemented numerically.

 \subsection*{Update step}
 
 Given the stress and deformation gradient from time $t$, $\sigma_{ij}^t$ and $F_{ij}^t$ respectively, as well as the new deformation gradient, $F_{ij}^{t+\Delta t}$, the goal is to obtain the updated stress $\sigma_{ij}^{t+\Delta t}$. Following \cite{dunatunga2015}, we begin by calculating the updated density $\rho^{t+\Delta t}=\rho^{t=0}/J$ where $J=\text{det}(F_{ij}^{t+\Delta t})$ is the Jacobian. In all our simulations, we begin by assuming $\rho^{t=0}=\rho_c$; because we gradually ramp up gravity, the system begins as a granular assembly barely in contact and at zero pressure.  If the deformed density is smaller than the critical density we set $\sigma_{ij}^{t+\Delta t}=0$ according to \textbf{Eq.S1}. Otherwise, we proceed to obtain updated stress $\sigma_{ij}^{t+\Delta t}$ using a variant of the radial return algorithm \cite{simo1985consistent} as we describe next.

 We start by assuming a purely elastic step, i.e. $D^p_{ij}=0$, under \textbf{Eq.S1}, which updates the stress to a ``trial stress" state ($\sigma_{ij}^{tr}$).  If the trial stress results in an equivalent shear stress $\bar{\tau}^{tr}=\sqrt{\sigma_{ij}^{tr'}\sigma_{ij}^{tr'}/2}$ that is less than $\mu_cP^{tr}$, it is accepted as the updated stress.  If not, it is then adjusted as shown in \textbf{Fig. S1}, which is described mathematically by
 
 \begin{equation}\tag{S2}
 	\sigma_{ij}^{t+\Delta t}	= \sigma_{ij}^{tr'}(\mu_cP^{t+\Delta t}/\bar{\tau}^{tr})-P^{t+\Delta t}\delta_{ij}.
 	\label{projection}
 \end{equation}
 In \textbf{Fig. S1}, $\sigma_{ij}^{t}$ is the stress state at the beginning of the step. Since $P^{tr}=P^{t+\Delta t}$ due to the isochoric plastic flow assumption, the effective shear stress $\bar{\tau}$ should reduce following a constant pressure route to reside on the yield surface and $\sigma_{ij}^{t+\Delta t}$ is updated accordingly. This essentially represents usage of a tangent modulus to return the trial stress state $\sigma_{ij}^{tr}$ back to the yield surface at the end of the increment \cite{simo1985consistent}.  
 
\renewcommand{\thefigure}{\textbf{S\arabic{figure}}}
\setcounter{figure}{0}

\begin{figure}
 	\centering
 	\epsfig{trim= 4in 1.2in 4in 1.2in, clip=true,file=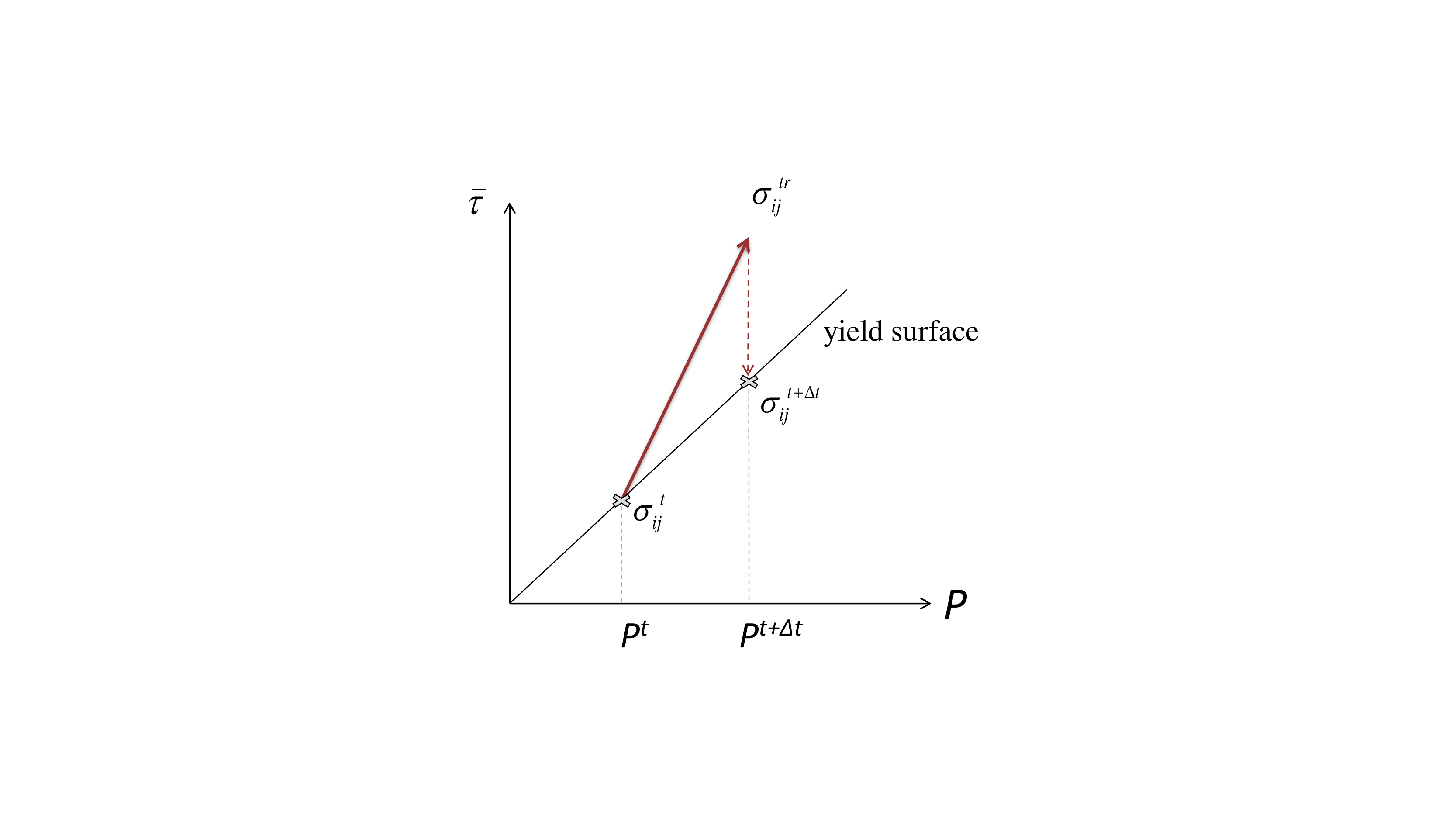,width=3in}

 \caption{\textbf{Update step.} Projection of the trial stress state to the yield surface per equation \textbf{Eq.S2}.}
  \end{figure}

 \subsection*{Encoding the material model and geometric inputs}
 The constitutive relation is implemented as a custom VUMAT subroutine in Abaqus finite element package to obtain numerical results from the theory. The meshed space in plane-strain condition represents a quasi-2D, i.e. one element thick, bed of material with in-plane nodal constraints to ensure plane-strain motion. We model the intruder as a rigid surface within the granular body that moves at an assigned, constant rate.  This choice relinquishes the need to define a complex contact routine between multiple objects, which would bring in additional contact parameters. This simplicity comes at a cost that the simulation conditions are not exactly the same as the experimental conditions used in the literature. We showed in the main paper that this assumption does not bring in noticeable error since quantitative agreement is established with experimental data. To obtain $\alpha_x^{\text{Theory}}$ and $\alpha_z^{\text{Theory}}$, same mesh has been used for a given attack angle ($\beta$=constant) and only the direction of the intrusion angle $\gamma$ has been adjusted in the FEM model to minimize mesh effects. 
 Similar approach is employed for 3D cases in terms of the boundary condition. The exact same mesh structure is used for horizontal and vertical V-shaped intruders to eliminate any attribute of mesh dependence.

 \subsection*{Simultaneous dense and disconnected matter}
 As an example, in the case of an flat plate traveling within granular material, one would expect that behind the intruder, the free space created by movement of the object would cause open states in its wake. Consequently the open material above the region should fall down and fill the cavities caused by the intruder movement as shown in density plot in \textbf{Fig. S2}. The open states in the back of the intruder are more or less persistent after reaching steady-state deformation while in the front there are isolated areas that show a limited amount of reduction in density. As expected, the figure shows these locations form in the trailing path of the intruder.

  	\begin{figure}
  		\epsfig{file=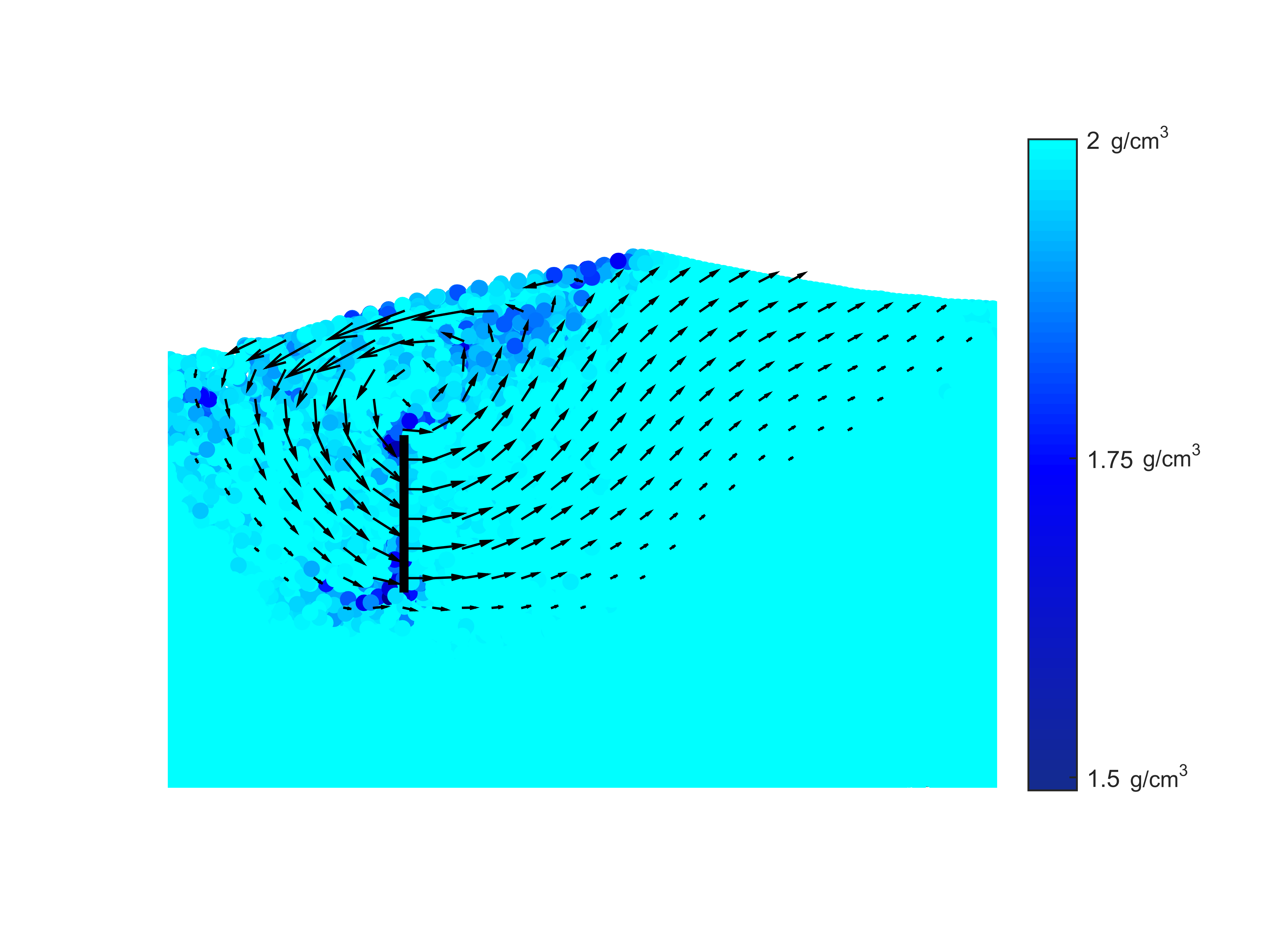,width=5in}

  	\caption{\textbf{Dense vs. open states.} Plot of the local density in the granular bed in the wake of a flat vertical intruder moving rightward using FEM. Densities below $\rho_c=\ $2 g/cm$^3$ denote disconnected material in 'open state'.}
   	\end{figure}

 \section*{S2. Resistive force theory for slender bodies}
 
 The general form of RFT was presented in the main paper for granular materials and viscous fluids. Here we describe how the analysis applies in the case when one of the dimensions is much longer than the others.  We consider a long horizontal rod of length $L$ and radius $a$ traveling within a horizontal plane at a speed $v$ angled $\phi$ away from the direction tangent to the bar as shown in \textbf{Fig. S3}.

  	\begin{figure}
  		\includegraphics[trim= 1in 7.5in 2.4in .7in, clip=true, width=6in]{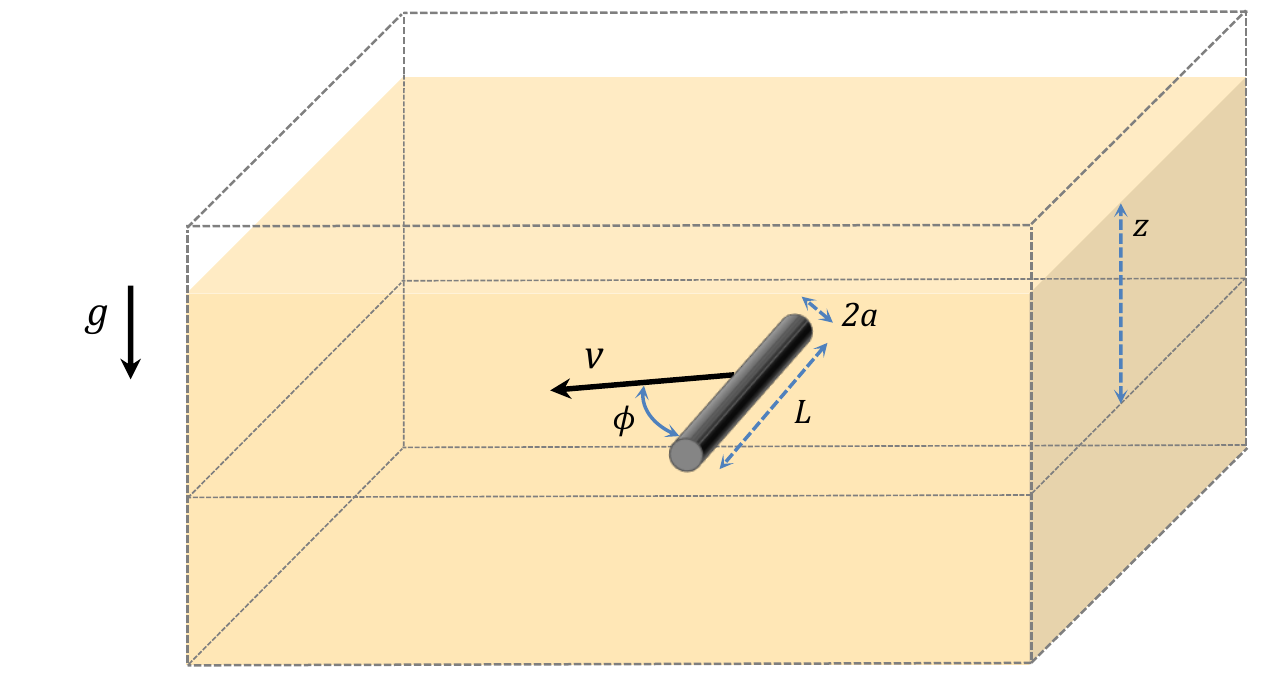}

  	\caption{\textbf{Slender body RFT.} schematic and geometry of a 1D object.}
  	
   	\end{figure}

 In the case of a granular material, the force on the rod can depend on $\phi$, $L$, $a$, $v$, the granular weight density, $\rho_c g$, the internal friction,  $\mu_c$, depth, $z$, and rod surface friction, $\mu_w$. Let the units of length, time, and mass be $[\textnormal{Length}]=z$, $[\textnormal{Time}]=z/v$ and $[\textnormal{Mass}]=\rho_c g z^4/v^2$ respectively.  Then the dimensionless force $\mathbf{F}/\rho_c g z^3$  is found as a function of the dimensionless inputs $\phi$, $\mu_c$, $\mu_w$, $L/z$, and $a/z$ as:
 \begin{equation}\tag{S3}
 	\mathbf{F_{\text{gran}}}= \rho_c g z^3\ \mathbf{\Psi}_{\text{gran}}(\phi,\mu_c,\mu_w, a/z,L/z).
 \end{equation}
 
 \noindent
 RFT superposition in the slender-body case is the notion that the force on any internal segment of the bar is well-approximated by the force on a similarly moving isolated bar of that length.  For the RFT superposition rule to be a strong approximation of the total force, thus, it must be possible for the force formula above to increase proportional to $L$ in the large $L$ limit.  The above formula permits this only if the function is asymptotically linear in its last input, yielding $\mathbf{F}_{\text{gran}}\sim \rho_c g z^2L \ \mathbf{\Psi}_{\text{gran}}(\phi,\mu_c,\mu_w, a/z)$.  Because $a/L\ll1$, we may expand in this variable.  The force must vanish when $a=0$,  so the result leads at the linear-order in $a/L$, implying
 \begin{equation}\tag{S4}
 	\mathbf{F_{\text{gran}}^{\text{RFT}}}\sim \rho_c g z L a\ \mathbf{\Psi}_{\text{gran}}(\phi,\mu_c,\mu_w).
 \end{equation}
 
 We compare the legitimacy of this formula with the analogous one we would obtain for a viscous fluid.  Here, the forces can depend only on $ \phi, v, L, a,$ and $\eta$. The units are selected as $[\textnormal{Length}]=L, [\textnormal{Time}]=L/v, [\textnormal{Mass}]=\eta L^2/v$ and the dimensionless force becomes $\mathbf{F}/\eta L v$.   Dimensional analysis gives $
 \mathbf{F}_{\text{visc}}= \eta L v\ \mathbf{\Psi}_{\text{visc}}(\phi,a/L) $.   In the large-$L$ limit, the RFT assumption of asymptotic linearity in $L$ requires the function lose dependence in the last input, giving
 \begin{equation}\tag{S5}
 	\mathbf{F_{\text{visc}}}\sim \eta L v\ \mathbf{\Psi}_{\text{visc}}(\phi).
 \end{equation}
 We note, for a contradiction, that this formula is non-physical --- the formula purports that the drag force is independent of the body's radius $a$, and since the drag force must vanish in the $a=0$ limit, it follows that $\mathbf{\Psi}_{\text{visc}}(\phi)=\mathbf{0}$ and the force should thus vanish for all $a$.  This contradiction means that the asymptotic linearity assumption needed for RFT cannot apply.  By contrast, the force formula in the plastic case, \textbf{Eq.S4}, shows a permissible dependence on $a$ and on depth $z$; in fact it is identical in form to experimental observations on slender-body motion in grains \cite{zhang2014effectiveness}.
 
 With the lack of a proper scaling in the viscous case, RFT's historical usage in fluids has required a further approximation, that of a problem-specific reference length-scale $\lambda$.  The resistive force per length is calculated for straight, slender bodies of length $\lambda$ and the result is \emph{imposed} as a local drag rule to find the forces on a possibly tortuous body of arc-length $L\gg \lambda$. The local drag coefficients achieved in this way maintain dependence on $\lambda/a$, which brings about a well-documented accrual of error in the RFT calculation the longer $L$ gets compared to $\lambda$ \cite{gray1955propulsion,lighthill1976flagellar}.

 \section*{S3. The influence of  additional granular constitutive phenomena on the RFT reduction}
 We briefly discuss how a more detailed granular constitutive model is likely to affect the collapse to RFT.  Rate-dependence in granular rheology arises through a rate-sensitive coefficient of friction $\mu_c(I)$ where the inertial number is $I=\dot{\gamma} d/\sqrt{P/\rho_s}$ with $\dot{\gamma}$ the shear-rate, $d$ the particle diameter, $P$ the pressure, and $\rho_s$ the solid grain density.  Generally speaking, when this number is larger than $\sim 0.1$ the the variation of $\mu_c$ with $I$ is non-negligible \cite{dacruz05}.  Defining units of length, time and mass as $L$, $L/v$ and $\rho_c g L^3/v^{2}$ respectively, we revisit our dimensional analysis of the garden hoe test and arrive at an additional dimensionless group $I_G=v^2 d^2/g L^2$ that accounts for inertial effects. Similarly, one could also include particle size-effects in the constitutive behavior, to account for the local strengthening of material that can occur if boundary features are small.  This could become non-negligible when $d/L$ is greater than $\sim 0.1$ \cite{kamrin12b,kamrin2015nonlocal}. Supposing size- and rate- dependent phenomena are included in the rheology and the problem is in a regime where either or both of these effects could matter, dimensional analysis of the garden hoe test instructs us to expect an extended general answer of the form
 
 \begin{equation}\label{hoe1}\tag{S6}
 	\mathbf{F}_{\text{gran}} = \rho_c g L^3 \mathbf{\Psi}_{\text{gran}}\left(\mu_c,\mu_w,b,\beta,\gamma,\frac{v^{2}d^{2}}{gL^2},\frac{d}{L}\right).
 \end{equation}
 Upon selecting a small reference length $\lambda$, the above formula yields an RFP relation $(\alpha_x, \alpha_z) =  2\mathbf{\Psi}(\beta,\gamma,\frac{v^2 d^2}{g\lambda^2},\frac{d}{\lambda})/\sin\beta$.  Superposing this result over the garden-hoe surface, the RFT result is
 
 \begin{equation}\label{hoe2}\tag{S7}
 	\mathbf{F}^{\text{RFT}}_{\text{gran}} = \rho_c g L^3 \mathbf{\Psi}_{\text{gran}}\left(\mu_c,\mu_w,b,\beta,\gamma,\frac{v^2 d^2}{g\lambda^2},\frac{d}{\lambda}\right).
 \end{equation}	
 The lingering dependence of the superposition results on $\lambda$ within the arguments of $\mathbf{\Psi}_{\text{gran}}$ could potentially cause the RFT prediction of (\ref{hoe2}) to differ from the exact answer in (\ref{hoe1}). Note that if $I_G$ is large enough to affect $\mathbf{\Psi}_{\text{gran}}$,  $\alpha_x$ and $\alpha_z$ will depend explicitly on the speed, contrary to the rate-independent RFP's that arise in the main text of the paper.

 \section*{S4. 3D superpositions tests in flat geometries}
 In the main paper, we evaluated the validity of the superposition principle in granular and viscous fluid for a V-shaped intruder. This geometry was chosen because it is applicable to many shapes found in nature given it's obtuse angle. In this section we study the most basic intrusion which is movement of flat plates. The procedure is exactly similar to what described in the main paper. \textbf{Fig. S4} shows the resistive forces found by superposition of two individual plates versus the actual resistive force for movement of a flat plat in horizontal orientation for viscous and granular case. The norm of the error in granular case is 6.5 \% which is similar to what was found for V-shape intruder. While the same geometry in a viscous fluid will show a norm of error about 42\%. The reason for an increase in the viscous case error for V-shape and flat intruder is in the lateral force component. Interestingly, the V-shape intruder seemed to be able to minimize the lateral force error because of the choice of the angle and therefore reducing the total norm of error to 36\%.

 	\begin{figure}
 		\includegraphics[trim= .1in .1in .1in .1in, clip=true, width=6in]{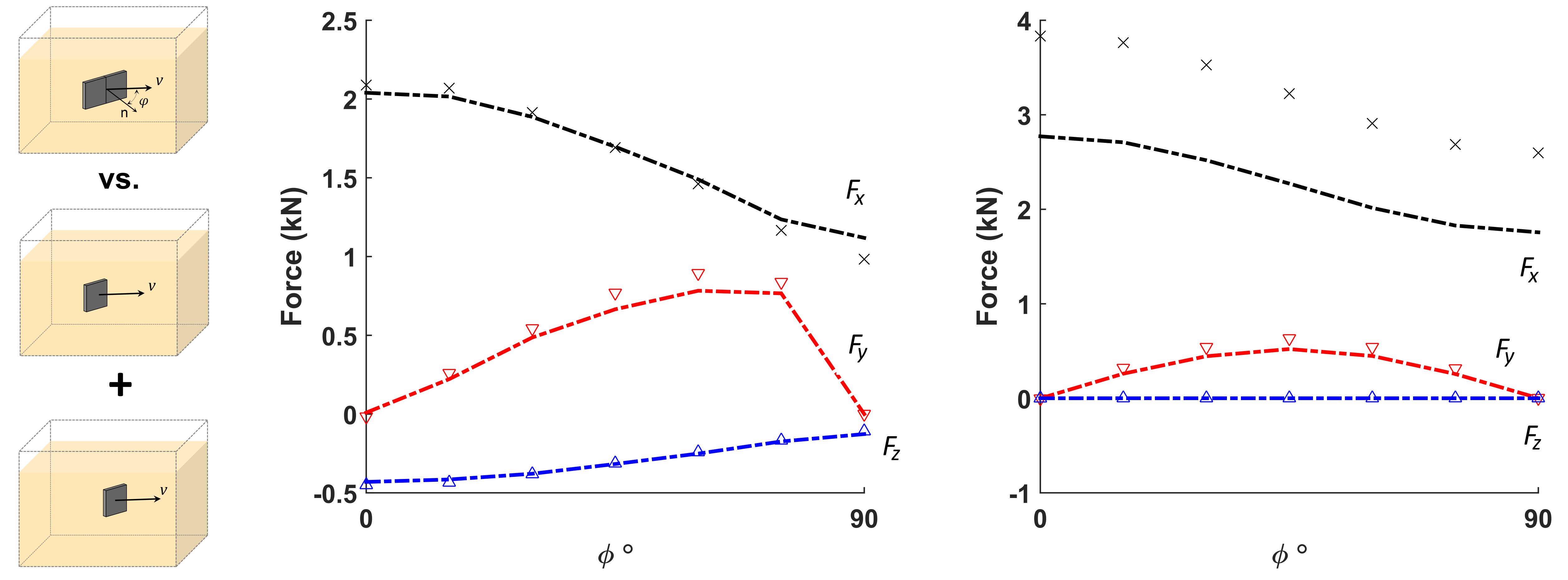}

 	\caption{\textbf{Evaluation of superposition for a flat intruder.}  The drag force, $F_x$, acting on a rightward moving submerged flat intruder  ($\times$),  and the components of lift, $F_z$ (\color{red} $\bigtriangledown$\color{black}), and lateral force, $F_y$ (\color{blue} $\triangle$\color{black}), compared with superposition values (dashed) at various orientation angles, $\varphi$, for horizontal intruder alignments in granular flow (left, $\rho_c=4 $g/cm$^3$, $\mu=$0.4, depth=0.6m) and viscous fluid (right, $\eta v$=15N/m). Intruder consists of two square plates of side length 20cm.}
  	\end{figure}

 \bibliography{scibib}

\begin{thebibliography}{10}

\bibitem{dickinson2000animals}
M.~H. Dickinson, {\it et~al.\/}, {\it Science\/} {\bf 288}, 100 (2000).

\bibitem{vogel1996life}
S.~Vogel, {\it Life in moving fluids: the physical biology of flow\/}
  (Princeton University Press, 1996).

\bibitem{lauder2002experimental}
G.~V. Lauder, J.~C. Nauen, E.~G. Drucker, {\it Integrative and Comparative
  Biology\/} {\bf 42}, 1009 (2002).

\bibitem{wang2005dissecting}
Z.~J. Wang, {\it Annu. Rev. Fluid Mech.\/} {\bf 37}, 183 (2005).

\bibitem{li2013terra}
C.~Li, T.~Zhang, D.~I. Goldman, {\it Science\/} {\bf 339}, 1408 (2013).

\bibitem{thorpe2007origin}
S.~K. Thorpe, R.~Holder, R.~Crompton, {\it Science\/} {\bf 316}, 1328 (2007).

\bibitem{biewener1990biomechanics}
A.~A. Biewener, {\it Science\/} {\bf 250}, 1097 (1990).

\bibitem{bhushan2009biomimetics}
B.~Bhushan, {\it Philosophical Transactions of the Royal Society of London A:
  Mathematical, Physical and Engineering Sciences\/} {\bf 367}, 1445 (2009).

\bibitem{ma2013controlled}
K.~Y. Ma, P.~Chirarattananon, S.~B. Fuller, R.~J. Wood, {\it Science\/} {\bf
  340}, 603 (2013).

\bibitem{williams2014self}
B.~J. Williams, S.~V. Anand, J.~Rajagopalan, M.~T.~A. Saif, {\it Nature
  communications\/} {\bf 5} (2014).

\bibitem{Ijspeert09032007}
A.~J. Ijspeert, A.~Crespi, D.~Ryczko, J.-M. Cabelguen, {\it Science\/} {\bf
  315}, 1416 (2007).

\bibitem{maladen2009undulatory}
R.~D. Maladen, Y.~Ding, C.~Li, D.~I. Goldman, {\it science\/} {\bf 325}, 314
  (2009).

\bibitem{bekker1960off}
M.~G. Bekker, {\it Research and development in Terramechanics\/}  (1960).

\bibitem{meirion2010empirical}
G.~Meirion-Griffith, M.~Spenko, {\it Aerospace Conference, 2010 IEEE\/} (IEEE,
  2010), pp. 1--6.

\bibitem{johnson2010measurement}
L.~Johnson, R.~King, {\it Journal of Terramechanics\/} {\bf 47}, 87 (2010).

\bibitem{wong2012predicting}
J.~Wong, {\it Journal of Terramechanics\/} {\bf 49}, 49 (2012).

\bibitem{uehara2003low}
J.~Uehara, M.~Ambroso, R.~Ojha, D.~Durian, {\it Physical review letters\/} {\bf
  90}, 194301 (2003).

\bibitem{shergold2004mechanisms}
O.~A. Shergold, N.~A. Fleck, {\it Proceedings of the Royal Society of London A:
  Mathematical, Physical and Engineering Sciences\/} (The Royal Society, 2004),
  vol. 460, pp. 3037--3058.

\bibitem{lighthill75}
J.~Lighthill, {\it Mathematica Biofluiddynamics\/} (SIAM, 1975).

\bibitem{ding2011drag}
Y.~Ding, N.~Gravish, D.~I. Goldman, {\it Physical Review Letters\/} {\bf 106},
  028001 (2011).

\bibitem{maladen2011RSociety}
R.~D. Maladen, Y.~Ding, P.~B. Umbanhowar, A.~Kamor, D.~I. Goldman, {\it Journal
  of The Royal Society Interface\/} {\bf 8}, 1332 (2011).

\bibitem{schofield1968critical}
A.~Schofield, P.~Wroth  (1968).

\bibitem{kamrin10a}
K.~Kamrin, {\it Int. J. Plasticity\/} {\bf 26}, 167 (2010).

\bibitem{kamrin12b}
K.~Kamrin, G.~Koval, {\it Phys. Rev. Lett.\/} {\bf 108}, 178301 (2012).

\bibitem{henann13}
D.~L. Henann, K.~Kamrin, {\it Proceedings of the National Academy of
  Sciences\/} {\bf 110}, 6730 (2013).

\bibitem{chen2013limit}
W.-F. Chen, {\it Limit analysis and soil plasticity\/} (Elsevier, 2013).

\bibitem{gray1955propulsion}
J.~Gray, G.~Hancock, {\it Journal of Experimental Biology\/} {\bf 32}, 802
  (1955).

\bibitem{lauga2009hydrodynamics}
E.~Lauga, T.~R. Powers, {\it Reports on Progress in Physics\/} {\bf 72}, 096601
  (2009).

\bibitem{rodenborn13}
B.~Rodenborn, C.-H. Chen, H.~L. Swinney, B.~Liu, H.~Zhang, {\it Proceedings of
  the National Academy of Sciences\/} {\bf 110}, E338 (2013).

\bibitem{dunatunga2015}
S.~Dunatunga, K.~Kamrin, {\it Journal of Fluid Mechanics\/} {\bf 779}, 483
  (2015).

\bibitem{hill1956new}
R.~Hill, {\it Journal of the Mechanics and Physics of Solids\/} {\bf 5}, 66
  (1956).

\bibitem{Abaqus}
Dassault Syst{\`e}mes Simulia, Providence, RI, {\it Abaqus Reference
  Manuals\/}, 6th edn. (2011).

\bibitem{goldman14}
D.~I. Goldman, {\it Reviews of Modern Physics\/} {\bf 86}, 943 (2014).

\bibitem{nedderman}
R.~M. Nedderman, {\it Statics and kinematics of granular materials\/}
  (Cambridge University Press, 2005).

\bibitem{dacruz05}
F.~{da Cruz}, S.~Emam, M.~Prochnow, J.-N. Roux, F.~Chevoir, {\it Physical
  Review E\/} {\bf 72}, 021309 (2005).

\bibitem{ding2012mechanics}
Y.~Ding, S.~S. Sharpe, A.~Masse, D.~I. Goldman  (2012).

\bibitem{simo1985consistent}
J.~C. Simo, R.~L. Taylor, {\it Computer methods in applied mechanics and
  engineering\/} {\bf 48}, 101 (1985).

\bibitem{zhang2014effectiveness}
T.~Zhang, D.~I. Goldman, {\it Physics of Fluids (1994-present)\/} {\bf 26},
  101308 (2014).

\bibitem{lighthill1976flagellar}
J.~Lighthill, {\it Siam Review\/} {\bf 18}, 161 (1976).

\bibitem{kamrin2015nonlocal}
K.~Kamrin, D.~L. Henann, {\it Soft matter\/} {\bf 11}, 179 (2015).

\end{thebibliography}
 
 \bibliographystyle{Science}

\end{document}